\documentclass[12pt]{article}

\usepackage{latexsym,amsmath,amscd,amssymb,framed,graphics,mathrsfs}
\usepackage{enumerate}

\usepackage{graphicx}
\usepackage{diagrams}

\usepackage[colorlinks]{hyperref}
\usepackage{url}
\usepackage{colordvi}
\usepackage{color}
\usepackage{bbm}
\usepackage{cancel}

\usepackage{soul}
\setstcolor{blue}
\usepackage{caption}

\usepackage[all]{xy}

\usepackage{frame}

\makeatletter

\@addtoreset{figure}{section}
\def\thefigure{\thesection.\@arabic\c@figure}
\def\fps@figure{h, t}
\@addtoreset{table}{bsection}
\def\thetable{\thesection.\@arabic\c@table}
\def\fps@table{h, t}
\@addtoreset{equation}{section}

\makeatother

\DeclareMathOperator{\diff}{d}
\newcommand{\pp}[2]{\frac{\partial #1}{\partial #2}} 
\newcommand{\dede}[2]{\frac{\delta \!#1}{\delta \!#2}}
\newcommand{\dd}[2]{\frac{\diff \!#1}{\diff \!#2}}

\DeclareMathOperator{\Hh}{\mathcal{H}}
\DeclareMathOperator{\Hph}{\mathcal{H}^\prime}
\DeclareMathOperator{\Ch}{\mathcal{C}}

\DeclareMathOperator{\Ll}{\mathcal{L}}
\DeclareMathOperator{\Al}{\mathcal{A}}
\DeclareMathOperator{\Ah}{\mathcal{A}}
\DeclareMathOperator{\Aph}{\mathcal{A}^\prime}
\DeclareMathOperator{\Bh}{\mathcal{B}}
\DeclareMathOperator{\Bph}{\mathcal{B}^\prime}

\DeclareMathOperator{\Hmh}{\mathsf{H}}
\DeclareMathOperator{\Amh}{\mathsf{A}}
\DeclareMathOperator{\Bmh}{\mathsf{B}}
\DeclareMathOperator{\Fmh}{\mathsf{F}}
\DeclareMathOperator{\Smh}{\mathsf{S}}
\DeclareMathOperator{\Cmh}{\mathsf{C}}

\DeclareMathOperator{\sigmafr}{\boldsymbol{\sigma} ^{\rm fr}}
\DeclareMathOperator{\sigmafrzero}{\boldsymbol{\sigma} ^{\rm fr(0)}}
\DeclareMathOperator{\ji}{\mathbf{j}_i}
\DeclareMathOperator{\js}{\mathbf{j}_s}

\DeclareMathOperator{\mv}{\mathbf{m}}
\DeclareMathOperator{\vv}{\mathbf{v}}
\DeclareMathOperator{\uv}{\mathbf{u}}
\DeclareMathOperator{\Qv}{\mathbf{Q}}
\DeclareMathOperator{\Fv}{\mathbf{F}}
\DeclareMathOperator{\Rv}{\mathbf{R}}
\DeclareMathOperator{\hatnv}{\mathbf{\hat{n}}}

\newtheorem{theorem}{Theorem}[section]

\newtheorem{remark}[theorem]{Remark}

\begin{document}
\title{Single and Double Generator Bracket Formulations of Geophysical Fluids with Irreversible Processes}

\author{
  Christopher Eldred\\
  \texttt{christopher.eldred@inria.fr}\\
Univ. Grenoble Alpes, Inria, CNRS, Grenoble INP, LJK  \and
  Fran\c{c}ois Gay-Balmaz\\
  \texttt{francois.gay-balmaz@lmd.ens.fr 
}\\
  CNRS, Ecole Normale Sup\'erieure de Paris, LMD
}

\maketitle

\abstract{
The equations of reversible (inviscid, adiabatic) fluid dynamics have a well-known variational formulation based on Hamilton's principle and the Lagrangian, to which is associated a Hamiltonian formulation that involves a Poisson bracket structure. These variational and bracket structures underlie many of the most basic principles that we know about geophysical fluid flows, such as conservation laws. However, real geophysical flows also include irreversible processes, such as viscous dissipation, heat conduction, diffusion and phase changes. Recent work has demonstrated that the variational formulation can be extended to include irreversible processes and non-equilibrium thermodynamics, through the new concept of thermodynamic displacement. By design, and in accordance with fundamental physical principles, the resulting equations automatically satisfy the first and second law of thermodynamics. Irreversible processes can also be incorporated into the bracket structure through the addition of a dissipation bracket. This gives what are known as the single and double generator bracket formulations, which are the natural generalizations of the Hamiltonian formulation to include irreversible dynamics. Here the variational formulation for irreversible processes is shown to underlie these bracket formulations for fully compressible, multicomponent, multiphase geophysical fluids with a single temperature and velocity. Many previous results in the literature are demonstrated to be special cases of this approach. Finally, some limitations of the current approach (especially with regards to precipitation and nonlocal processes such as convection) are discussed, and future directions of research to overcome them are outlined.}


\section{Introduction}

Variational formulations based on Hamilton's principle and the Lagrangian for geophysical fluids are well established for reversible dynamics \cite{HoMaRa2002,Sa1983,Sa1988}. Through the Legendre transform (when it is invertible), there is an associated Hamiltonian formulation based on Lie-Poisson brackets \cite{HoMaRaWe1985,MaWe1983,MoGr1980,Shepherd1990}. Such approaches have proven to be a powerful tool  for the derivation of various models and consistent approximations in geophysical fluid dynamics, see, e.g., \cite{Cr1991,DeSa2005,Dubos2014,Ho1996,MiSa1985,Ol2006,Ri1981,Sa1983,Sa1985,Shepherd1993,Tort2014a}. More recently, these methods have been strongly exploited for the development of numerical schemes, both on the Lagrangian (variational) side \cite{BaGB2018,BrBaBiGBML2018,DeGaGBZe2014,PaMuToKaMaDe2010} and on the Hamiltonian (bracket) side \cite{Dubos2015,Eldred2017,EldredDubos2018,Gassmann2013,Gassmann2008,Salmon2004,Tort2015}, to name a few.

However, until recently such formulations were restricted to purely reversible processes. A more complete picture of geophysical fluids includes irreversible processes, such as viscous dissipation, heat conduction, diffusion and phase changes. An extension of the variational approach for geophysical fluids to include irreversible processes was developed in  \cite{GayBalmaz2017}, which is based on the general variational formulation of non-equilibrium thermodynamics in \cite{GBYo2017a,GBYo2017b}. Unlike previous approaches, this is a systematic construction for irreversible processes that requires only an expression for the entropy generation rate of the irreversible process. The key idea is the concept of thermodynamic displacement, and by design, the resulting equations satisfy both the first and second laws of thermodynamics. These are two fundamental principles of nature that are believed to hold for a wide range of physical processes. 

From the bracket perspective, a natural generalization to include irreversible processes is the combination of a Poisson bracket for the reversible dynamics with a dissipation bracket for the irreversible dynamics. The inclusion of dissipative or irreversible phenomena in Hamilton's equations through a modification of the Poisson bracket has been initiated by \cite{Gr1984,Ka1984,Mo1984a,Mo1984b}. This has largely followed two approaches: the single generator formulation \cite{EdBe1991b,EdBe1991a} and the double generator formulation \cite{GrOtt1997,Ka1984,Mo1986,OtGr1997}. See Section \ref{bracket-formulations} for further discussion of these. These two approaches differ only in how the dissipation bracket is specified, and will of course give the same equations of motion.  Single generator brackets for a variety of Newtonian and non-Newtonian fluids are discussed in \cite{BeEd1994}. A specific variant of the double generator approach is a metriplectic system \cite{Mo1986}, which places even stronger constraints on the dissipative bracket. An example of a metriplectic system based on the idea of Casimir decay is found in \cite{GBHo2013,GBHo2014}. The most prominent example of the double generator formalism is the general equation for the non-equilibrium reversible-irreversible coupling (GENERIC) approach \cite{GrOtt1997,OtGr1997}, which is actually metriplectic. The majority of the work for fluids using GENERIC (and in the single generator bracket formulations as well) has been done in the fields of complex fluids, such as polymer melts, liquid crystals and other non-Newtonian fluids. In contrast, here we are interested in the dynamics of Newtonian fluids with multiple components and phases dominated by the influence of gravity and rotation; these are known as geophysical fluids.

The construction of the single and double generator formulations as well as the relation between them has been most of the time very empirical, and there is a lack of a general constructive procedure able to directly produce these brackets from first principles. In this paper, we show that the variational formulation for geophysical fluids with irreversible processes \cite{GayBalmaz2017} underlies both single and double generator bracket formulations, in the context of a multicomponent, multiphase fully compressible fluid undergoing reversible and irreversible processes (viscous dissipation, heat conduction, diffusion and phase changes). Starting from the variational formulation for a given arbitrary Lagrangian and taking the Legendre transform of the resulting system in order to express it in momentum variables, we present a systematic construction for the single generator bracket and the double generator bracket, for general expressions of the thermodynamic fluxes in terms of the thermodynamic forces. We also present a systematic construction of a metriplectic (or GENERIC) bracket when the thermodynamic fluxes depend linearly on the thermodynamic forces. These formulations are shown to reduce to existing, known bracket formulations (e.g., \cite{BeEd1994,Mo1984b}) for classical hydrodynamics in the case of a single component. Another advantage of the variational formulation is that it has both a Lagrangian and an Eulerian version that are systematically related through a reduction of the variational principle by the relabelling symmetry. This property immediately transfers to the bracket side and thus yields both a Lagrangian and Eulerian version of the single, double, and metriplectic brackets. In particular, in the Lagrangian version, the reversible dynamics is governed by canonical Poisson brackets. We shall however only focus on the Eulerian formulation on the bracket side in this paper. The thermodynamic fluxes can be interpreted as either physical irreversible processes, or as subgrid turbulent processes. If using the latter, the resulting formulation has strong similarities to the approach in \cite{Gassmann2015} for the development of parameterizations that are consistent with the second law of thermodynamic, and we will show that the formulation in \cite{Gassmann2015} is a specific example of our general framework. In particular it is a choice of parameterization of the thermodynamic fluxes in terms of the thermodynamic forces.

In deriving the variational formulation, there are three key assumptions that are made:
\begin{enumerate}
\item Local thermodynamic equilibrium: at a given point in space and time, for some local neighborhood, thermodynamic equilibrium holds and the state can be described by some set of intensive thermodynamic variables. This means, for example, it is meaningful to speak of the temperature and pressure of the fluids.
\item The domain $\Omega$ is materially closed and has rigid boundaries. We assume $\uv \cdot \hatnv = 0$ on $\partial \Omega$ for the reversible dynamics, and $\uv|_{\partial \Omega} = 0$ for the irreversible dynamics.
\item All components of the fluid have the same temperature $T$, and move at the same (barycentric) velocity $\uv$.
\end{enumerate}
The first approximation is a common assumption of continuum mechanics, and is believed to hold for geophysical fluids of sufficiently high density. For example, it holds in the Earth's atmosphere below altitudes of approximately 80km. The second assumption is a fairly standard one in geophysical fluid dynamics, although it is insufficient to describe the interaction between the various components of the climate system (atmosphere, ocean, land), and also the possibility of a pressure boundary condition at the top of the atmosphere. The third assumption is known to be invalid for fluids containing larger hydrometeors (especially ice) and in the presence of precipitation \cite{Bannon2002}, although it is commonly made. The last two approximations will be removed in future work.

Before proceeding further, it is useful to discuss briefly the meaning of the terms reversible and irreversible. As an extensive quantity, the rate of change of entropy in some volume is given by ${\rm d}s = {\rm d}s_i+{\rm d}s_e$, where ${\rm d}s_i$ is the production within the volume and ${\rm d}s_e$ is the flux of entropy across the boundaries. The second law of thermodynamics states that ${\rm d}s_i \geq 0$. A reversible processes has ${\rm d}s_i = 0$, while an irreversible process has ${\rm d}s_i > 0$. An adiabatically closed system has ${\rm d}s_e = 0$, and an isentropic process has ${\rm d}s = 0$. In an adiabatically closed system, reversible = isentropic, but this is not true in general. For example, the well-developed atmospheric boundary layer is isentropic, but not reversible. Instead, the local generation of entropy through irreversible processes is balanced by the transport of entropy. In this paper we will use only the distinction between reversible and irreversible processes.

\begin{figure}[h]
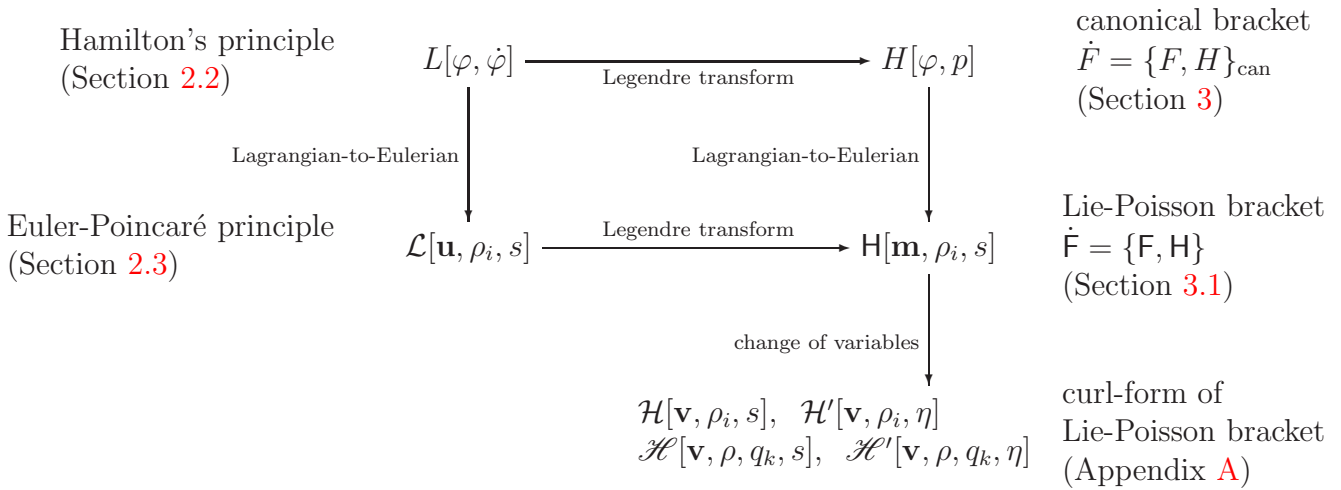

\begin{diagram}
\hspace{-0.2cm}
\begin{array}{l}
\text{Hamilton's principle}\\
\text{(Section \ref{lagrangian-formulation})}
\end{array}
 & L[\varphi, \dot\varphi] & \rTo_{\tiny  \text{Legendre transform}} &  H[\varphi, p] & \begin{array}{l} 
\text{canonical bracket}\\
\dot{F}=\{F,H\}_{\rm can}\\
\text{(Section \ref{hamiltonian-approach})}
\end{array}
\\
 & \dTo^{\tiny \text{Lagrangian-to-Eulerian}} & & \dTo^{\tiny \text{Lagrangian-to-Eulerian}} & \\
 \hspace{-0.9cm}
\begin{array}{l}
\text{Euler-Poincar\'e principle}\\
\text{(Section \ref{variational-reversible-eulerian})}
\end{array} & \Ll[\uv,\rho_i,s] & \rTo^{\tiny \text{Legendre transform}}& \mathsf{H}[\mathbf{m},\rho_i,s] & \begin{array}{l} 
 \text{Lie-Poisson bracket} \\
  \dot{\mathsf{F}}=\{\mathsf{F},\mathsf{H}\}\\
  \text{(Section \ref{lie-poisson-formulation})}
 \end{array} \\
 & & & \dTo^{\tiny \text{change of variables}} & \\
 & & & \hspace{-2.5cm}\begin{array}{ll}\mathcal{H}[\mathbf{v},\rho_i,s],\;\; \mathcal{H}'[\mathbf{v},\rho_i,\eta]\\\mathscr{H}[\mathbf{v},\rho, q_k,s],\;\;\mathscr{H}'[\mathbf{v},\rho, q_k,\eta]\end{array} & \begin{array}{l}
\text{curl-form of}\\
\text{Lie-Poisson bracket}\\
\text{(Appendix \ref{alternative-prognostic})}
\end{array}\\
\end{diagram}

\caption{A high level overview of the variational and Hamiltonian approaches to reversible dynamics, and the relationships between them. Both approaches start with a Lagrangian $L[\varphi, \dot\varphi]$ and its Eulerian version $\Ll[\uv,\rho_i,s]$ that characterize the fluid (it is also possible to use $\Ll[\uv,\rho_i,\eta]$, but this is not explored further here). The variational formulation then uses Hamilton's principle for $L$, that induces the Euler-Poincar\'e variational principle for $\mathcal{L}$, to derive the equations of motion. The Hamiltonian formulation instead uses a Legendre transform to obtain the Hamiltonian $\Hmh[\mv,\rho_i,s]$ from $\Ll[\uv,\rho_i,s]$. This is only possible when the Legendre transform is regular (invertible), which is usually but not always the case (for example, hydrostatic fluids in Eulerian coordinates have a irregular, non-invertible Legendre transform). The associated Poisson bracket is a Lie-Poisson bracket. To obtain the curl-form Poisson brackets, a change of variables is made from $(\mv,\rho_i,s)$ to $(\vv,\rho_i,s)$ (as explored in Appendix \ref{alternative-prognostic}). It is also possible to make further changes of variables, for example replacing $s$ with $\eta$, which is also explored in Appendix \ref{alternative-prognostic}.}
\label{reversible-overview-fig}
\end{figure}

\begin{figure}[h]
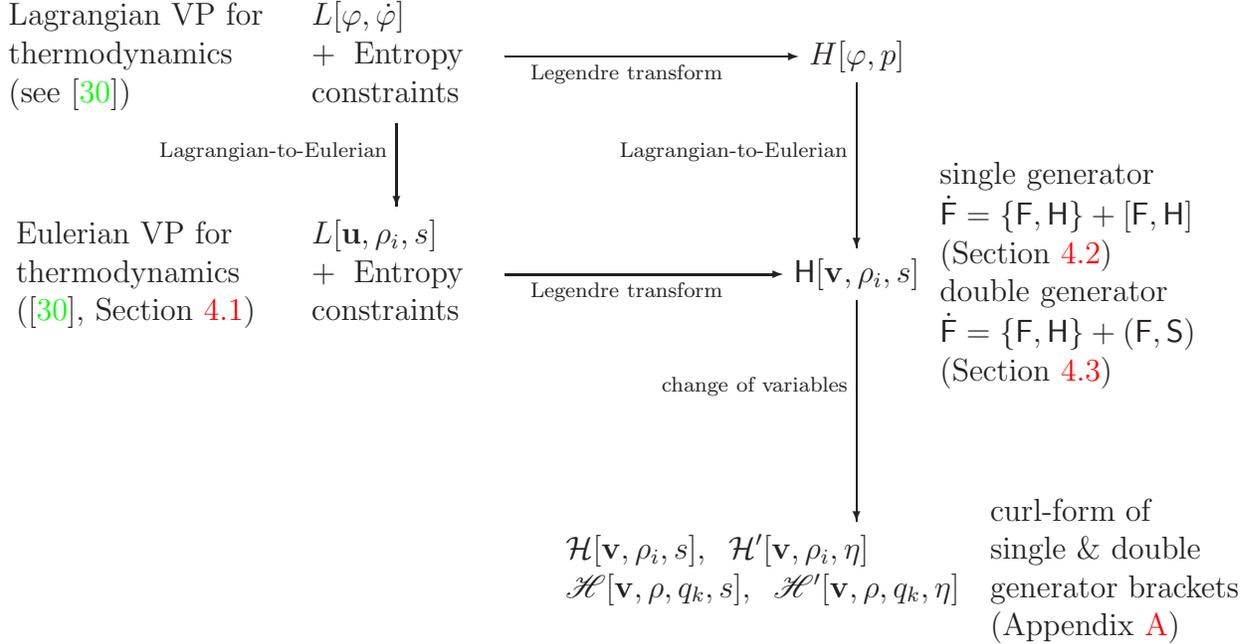

\begin{diagram}
\begin{array}{l}
\text{Lagrangian VP for}\\
\text{thermodynamics}\\
\text{(see \cite{GayBalmaz2017})}
\end{array} & \begin{array}{l}
L[\varphi, \dot\varphi]\\
+\;\;\text{Entropy}\\
\text{constraints}
\end{array} \;\;&\rTo_{\tiny \text{Legendre transform}}&  H[\varphi, p]\\
 & \dTo^{\tiny \text{Lagrangian-to-Eulerian}} &  & \dTo^{\tiny \text{Lagrangian-to-Eulerian}} & \\
 \begin{array}{l}
\text{Eulerian VP for}\\
\text{thermodynamics}\\
\text{(\cite{GayBalmaz2017}, Section \ref{variational-irr-section})}\\
\end{array} & \begin{array}{l}
L[\mathbf{u},\rho_i,s]\\
+\;\;\text{Entropy}\\
\text{constraints}
\end{array} \;\;&\rTo_{\tiny \text{Legendre transform}}&  \Hmh[\mathbf{v},\rho_i,s] & \hspace{-1.25cm}\begin{array}{l}
\text{single generator}\\
\dot{\mathsf{F}}=\{\mathsf{F},\mathsf{H}\}+[\mathsf{F},\mathsf{H}]\\
\text{(Section \ref{single-gen-bracket-section})}\\
\text{double generator}\\
\dot{\mathsf{F}}=\{\mathsf{F},\mathsf{H}\}+(\mathsf{F},\mathsf{S})\\
\text{(Section \ref{double-gen-bracket-section})}
\end{array}\\
 &  &  & \dTo(1,2)^{\tiny \text{change of variables}} & \\
  &  &  &  & \\
 &  &  & \hspace{-2.5cm} \begin{array}{ll}\Hh[\vv,\rho_i,s], \;\;\Hh^\prime[\vv,\rho_i,\eta]\\\mathscr{H}[\vv,\rho, q_k,s],\;\;\mathscr{H}'[\vv,\rho, q_k,\eta]\end{array} & \begin{array}{l}
\text{curl-form of}\\
\text{single \& double}\\
\text{generator brackets}\\
 \text{(Appendix \ref{alternative-prognostic})}
\end{array}
\end{diagram}
\caption{As Figure \ref{reversible-overview-fig}, but for the irreversible dynamics. Starting now from both a Lagrangian $\Ll[\uv,\rho_i,s]$ and entropy generation rates for the irreversible processes (included as a constraint), the variational formulation is obtained by using the variational principle developed in \cite{GayBalmaz2017,GBYo2017a,GBYo2017b}. The reversible part of the single and double generator bracket formulations are again obtained by taking the Legendre transform and yield the Lie-Poisson bracket. The dissipation bracket for both cases is obtained using the entropy generation rates, as discussed in Section \ref{bracket-formulations}. A change of variables from $(\mv,\rho_i,s)$ to $(\vv,\rho_i,s)$ again yields the curl-form, with a curl-form Poisson bracket and a curl-form dissipation bracket (see Appendix \ref{alternative-prognostic})}
\label{irreversible-overview-fig}
\end{figure}

The remainder of this paper is structured as follows. Sections \ref{lagrangian-section} and \ref{hamiltonian-approach} review (and extend to multicomponent fluids as needed) the well-known variational and Hamiltonian formulations for the reversible dynamics. Section \ref{bracket-formulations} then introduces irreversible dynamics in the variational formulation (following \cite{GayBalmaz2017}) and develops the associated single and double generator bracket formulations as well as the metriplectic formulation, including the reduction to single bracket generator of \cite{BeEd1994} and to the metriplectic formulation of \cite{Mo1984b} for a single component. The parameterization of thermodynamic fluxes in terms of thermodynamic forces to ensure the production of entropy is discussed in Section \ref{parameterizing-fluxes}, and two different variants are discussed: one based on an interpretation of thermodynamic fluxes as physical molecular-scale irreversible processes, and one treating them as subgrid turbulence fluxes. It is shown that the latter yields \cite{Gassmann2015}. Finally, Section \ref{conclusions} draws some conclusions and offers future directions of research. Appendix \ref{alternative-prognostic} discusses some alternative choices of prognostic variables that give rise to what are known as curl-form formulations and Appendix \ref{pv-kelvin-circulation} gives the Kelvin Circulation Theorem and the potential vorticity dynamics for the multicomponent, multiphase equations. A high-level overview of the various formulations and choice of predicted variables can be found in Figure \ref{reversible-overview-fig} for the reversible dynamics, and in Figure \ref{irreversible-overview-fig} for the irreversible dynamics.

\section{Variational Formulation for Reversible Dynamics}
\label{lagrangian-section}

Before proceeding with the introduction of irreversible processes, we first review in this section the variational formulation of multicomponent, multiphase compressible fluids undergoing reversible processes in both the Lagrangian (or material) and Eulerian (or spatial) descriptions. In the Lagrangian description, the variational principle is just the classical Hamilton's principle $\delta \int_0^TL[\varphi, \dot \varphi]dt=0$, applied to the fluid flow $\varphi$, where $L$ is the Lagrangian function of the fluid. The variational principle in the Eulerian description is then deduced from it.

%\begin{remark}\label{remark_Lagrangian}{\rm We would like to warn the reader that the word \textit{Lagrangian} is used here, and throughout the paper, for two different meanings. First there is the \textit{Lagrangian description}, also called \textit{material description} which refer to the description of fluid dynamics made by an observer following an individual fluid parcel as it moves through space and time. Such a description can be made for the Lagrangian and Hamiltonian formulations. Second, there is the \textit{Lagrangian formulation} (also referred to as the \textit{variational formulation}), which refers to a formulation expressed in terms of a Lagrangian function.}
%\end{remark}

%\todo{CE: I removed all the references to Langrangian formulation and replaced them with variational formulation, so it is not clear that this remark is needed anymore? Although perhaps the distinction between Lagrangian description and Lagrangian function should be emphasized?}

\subsection{Domain, Coordinate System and Prognostic Variables}

\paragraph{Domain and Coordinates.} We consider the reversible dynamics of a fully compressible, multicomponent, multiphase fluid in an arbitrary geopotential with an arbitrary equation of state, in a fixed domain $\Omega$ that is a closed subset of $\mathbb{R}^3$ with smooth boundary $\partial\Omega$. The coordinate system is rotating with rotation vector $\bf \Omega$, which is almost always chosen to correspond with the rotation of the underlying planetary body. Associated with the rotation is solid-body velocity of the coordinate system $\Rv$, which satisfies $\nabla \times \Rv = 2 \bf \Omega$\footnote{Note the gauge freedom here in the definition of $\Rv$. In $\mathbb{R}^3$ the standard choice is $\Rv = \bf \Omega \times \mathbf{r}$, where $\mathbf{r}$ is the position vector}. All the developments made in this paper generalize to the case when the fluid domain $\Omega$ is a Riemannian manifold, possibly with boundary. While we focus on the case $\Omega\subset \mathbb{R}^3$, we shall occasionally indicate how our formulation can be adapted to the general case. For instance, the relation between $\mathbf{R}$ and $\bf\Omega$ reads $\mathrm{d}\mathbf{R}^\flat=2 \bf \Omega$, where $\mathrm{d}$ is the exterior derivative and $\mathbf{R}^\flat$ is the one-form associated to the vector field $\mathbf{R}$ via the Riemannian metric.

\paragraph{Variables.}
The fluid is assumed to consist of $n$ components at local thermodynamic equilibrium, with a single temperature $T$ and relative (barycentric) velocity $\uv$. There is a dominant component (typically dry air or liquid water) along with $n-1$ sparse components. For example, seawater consists of two components: liquid water (dominant) and salt (sparse). Warm moist air consists of two components: dry air (dominant) and water substance (sparse, with liquid and vapor grouped together and assumed to satisfy the Clausius-Clapeyron equation). More complicated multicomponent fluids can be treated the same way. The assumption of a single temperature and velocity, although commonly made, is only somewhat justified in the atmosphere, especially when larger hydrometers are present \cite{Bannon2002}. Allowing different components to have different velocities is also required to treat precipitation. However, an extension to multiple interacting, open subsystems with distinct velocities and temperatures is deferred to future work. The fundamental variables are the mass densities of each component $\rho_i$, with $i\in\{1,\dots,n\}$, the specific entropy $\eta$ and the relative velocity $\uv$. In the case of reversible dynamics, we shall assume that there is no flux through the boundary, ie $\uv \cdot \hatnv = 0$, for $\hatnv$ the unit normal to $\partial \Omega$. Other important variables are the total mass density $\rho = \sum_i \rho_i$, specific volume $\alpha = \frac{1}{\rho}$, specific concentration $q_i = \frac{\rho_i}{\rho}$ of component $i$, and entropy density $s = \rho \eta$. There are also the absolute velocity $\vv$ and the absolute momentum density $\mv$, defined as
\[
\mv = \rho \vv = \rho (\uv + \Rv) \quad\quad  \vv = \uv + \Rv.
\]
We shall see later in Section \ref{lagrangian-formulation} how these definitions can be made in general for any Lagrangian.

For each of the $n$ components, we can predict either component density $\rho_i$ or specific concentration $q_i$, with the caveat that there must be at least one density predicted amongst the $n$. These are the $n$ mass variables. It is also possible to replace one of the $\rho_i$ or $q_i$ with total density $\rho$, typically it is the density $\rho_d$ of the dominant component that is replaced. For the mass variables, two choices are commonly made: the $n$ component densities $\rho_i$, or the total density plus $n-1$ sparse concentrations $(\rho,q_k)$, where $k\in \{1,\dots,n-1\}$ ranges over the sparse components. We must also predict an entropy variable chosen from the set $(\eta,s)$ and a velocity variable chosen from the set $(\uv,\vv,\mv)$, giving finally $n + 2$ prognostic variables. In the main text of this paper, we will choose as prognostic variables $(\uv,\rho_i,s)$ in the Lagrangian variational formulation, and $(\mv,\rho_i,s)$ in the Hamiltonian (Lie-Poisson) formulation. A discussion of the alternative sets $(\vv,\rho_i,s)$, $(\vv,\rho_i,\eta)$, $(\vv,\rho,q_k,\eta)$, and $(\vv,\rho,q_k,s)$, which are associated with curl-form Hamiltonian formulation, is given in Appendix \ref{alternative-prognostic}.

%\subsection{Vector Invariant Form}
%From standard texts in geophysical fluid dynamics (REFS), the equations of motion for a multicomponent fluid can be written as 
%\begin{eqnarray}
%\label{rhon-eqn}
%\partial_t \rho_i + \nabla \cdot (\rho_i \uv) = 0 \\
%\label{v-eqn}
%\partial_t \vv + \nabla \times \vv \times \uv + \nabla K + \nabla \Phi + \alpha \nabla p = 0 \\
%\label{m-eqn}
%\partial_t \mv + ADD THIS + \rho \nabla \Phi + \nabla p= 0 \\
%\label{S-eqn}
%\partial_t S + \nabla \cdot (S \uv) = 0 \\
%\label{s-eqn}
%\partial_t s + \uv \cdot \nabla s = 0
%\end{eqnarray}
%where $K = \frac{\uv \cdot \uv}{2}$ is the kinetic energy, $\Phi$ is the geopotential and $p = p(\alpha,s,q_i) = p(\frac{1}{\rho},\frac{S}{\rho},\frac{\rho_i}{\rho})$ is the pressure, whose specific expression comes from the equation of state. Only one of (\ref{S-eqn}) and (\ref{s-eqn}) is needed. Note that $\nabla \times \vv \times \uv = \nabla \times \uv \times \uv + 2 \bf \Omega \times \uv$ and $\partial_t \vv = \partial_t \uv$, which is a more common form.

\paragraph{Thermodynamics.} We assume that the internal energy $U$ can be characterized by the set of state variables as $U = U(\alpha,\eta,q_i)$. Therefore, the fundamental thermodynamic relationship can be expressed as
\[
{\rm d}U = -p {\rm d}\alpha + T {\rm d}\eta + \sum_i \mu_i {\rm d}q_i
\]
with pressure $p = -\pp{U}{\alpha}$, temperature $T = \pp{U}{\eta}$, and chemical potentials $\mu_i = \pp{U}{q_i}$. The chemical potentials are simply the partial Gibbs free energy and satisfy $\sum_i \mu_i = G = U + p \alpha - Ts$. It will also be useful to introduce the Gibbs-Duhem equation
\begin{equation}
\label{gibbs-duhem}
\sum_i q _i {\rm d} \mu_i + \eta {\rm d} T - \alpha {\rm d} p = 0.
\end{equation}

Note that here we work with the mass concentration $q _i$ (as typically done for geophysical fluids), not the molar concentration $n_i = \frac{q _i}{m_i}$, where $m_i$ is the molecular weight of component $i$. The molar concentration is more commonly used in physics, and would lead to a replacement of $\mu_i$ with $\frac{\mu_i^\prime}{m_i}$ where $\mu_i^\prime = \pp{U}{n_i}$, for the internal energy written as a function $U = U(\alpha,\eta,n_i)$.

\subsection{Hamilton's Variational Principle for Fluid motion}
\label{lagrangian-formulation}
In absence of irreversible processes, the equations of motion for fluid dynamics can be derived by applying Hamilton's variational principle to the Lagrangian function of the fluid. This is in agreement with a fundamental fact from classical reversible mechanics, namely that the motion of the mechanical system is governed by the Euler-Lagrange equations which, in turn, describe the critical points of the action functional of this Lagrangian among all possible trajectories with prescribed values at the temporal extremities. Hamilton's principle for fluid mechanics in the Lagrangian description has been discussed at least since the works of \cite{He1955}, for an incompressible fluid and  \cite{Ec1960,Se1959} for compressible flows. The independent variable in the Lagrangian description is uniquely the fluid flow $\varphi$, assigning the current positions $x=\varphi(t,X)\in\Omega$ at time $t$, of the fluid particles labelled by $X\in \Omega$. It is written as
\begin{equation}\label{HP}
\delta\int_0^TL[\varphi, \dot\varphi]{\rm d}t=0,
\end{equation}
for arbitrary variations $\delta\varphi$, vanishing at $t=0,T$, where $L=L[\varphi, \dot\varphi]$ is the Lagrangian of the fluid.
In the Lagrangian description, the mass densities and the entropy density, denoted $\varrho_i(X)$ and $S(X)$, are time independent, as a consequence of their conservation, hence they are not explicitly involved in the variational principle, although the Lagrangian depends parametrically on them.

While in the Lagrangian description this principle is a straightforward extension of the Hamilton principle of particles mechanics, in the Eulerian description the variational principle is much more involved and several approaches have been developed, see \cite{Br1970,Li1963,SeWh1968}. We refer to \cite{Sa1983,Sa1988} for further developments in the context of geophysical fluids. In \cite{HoMaRa2002}, the variational principle in Eulerian description is systematically obtained via the Euler-Poincar\'e reduction theory for several geophysical fluid models, by exploiting the relabelling symmetries. This is the point of view that we recall below.

\subsection{Variational Formulation in the Eulerian Description}
\label{variational-reversible-eulerian}
The Eulerian variables $\mathbf{u}$, $\rho_i$, $s$ are connected to their Lagrangian counterpart $\dot\varphi$, $\varrho_i$, $S$ as
\begin{align}
\dot\varphi(t,X)&=\mathbf{u}(t,\varphi(t,X))\label{EL_relation_u}\\
\varrho_i(X)&= \rho_i(t, \varphi(t,X))|\nabla\varphi(t,X)|\label{EL_relation_rho}\\
S(X)&= s(t, \varphi(t,X))|\nabla\varphi(t,X)|,\label{EL_relation_s}
\end{align}
where $|\nabla\varphi|$ denotes the Jacobian of the fluid flow. These formula are fundamental for the determination of the variational principle in the Eulerian description, deduced from the Hamilton principle.

\paragraph{Kinematic Equations.} In the Eulerian description, relations \eqref{EL_relation_rho} and \eqref{EL_relation_s} yield the kinematic equations for the mass densities $\rho_i$ and entropy density $s$ as the familiar conservation laws
\begin{align}
&\partial _t\rho_i + \nabla \cdot (\rho_i \,\mathbf{u}) =0\label{rho-kinematic}\\
&\partial_t s + \nabla \cdot (s\, \mathbf{u})=0.\label{S-kinematic}
\end{align}

\paragraph{Eulerian Variational Principle.} From the relabelling symmetries of fluid dynamics, the Lagrangian $L[\varphi, \dot\varphi]$ can be expressed in terms of the Eulerian fields $\mathbf{u}$, $\rho_i$, $s$ and thus defines the Lagrangian $\mathcal{L}$ in Eulerian variables as $L[\varphi, \dot\varphi]=\mathcal{L}[\mathbf{u},\rho_i, s]$, where \eqref{EL_relation_u}--\eqref{EL_relation_s} holds. Then, in the Eulerian description, Hamilton's principle \eqref{HP} yields, using \eqref{EL_relation_u}--\eqref{EL_relation_s} again, the variational principle of Euler-Poincar\'e type, \cite{HoMaRa2002},
\begin{equation}\label{EP_VP}
\delta\int_0^T\mathcal{L}[\mathbf{u},\rho_i, s]{\rm d}t=0,
\end{equation}
for constrained variations of the form
\begin{equation}\label{EP_variations} 
\delta \mathbf{u} =\partial _t \boldsymbol{\zeta} + \mathbf{u} \cdot \nabla \boldsymbol{\zeta} - \boldsymbol{\zeta}\cdot \nabla \mathbf{u} ,\;\;\;\; \delta \rho_i =- \nabla\cdot ( \rho_i \,\boldsymbol{\zeta} ),\;\;\;\; \delta s =- \nabla\cdot ( s \,\boldsymbol{\zeta} ),
\end{equation} 
where $\boldsymbol{\zeta}$ is a vector field with boundary condition $\boldsymbol{\zeta}\cdot\hat{\mathbf{n}}=0$ on $\partial\Omega$ and with $\boldsymbol{\zeta}=0$ for $t=0,T$. This vector field is connected to the variation of the fluid flow as
\[
\delta\varphi(t,X)=\boldsymbol{\zeta}(t,\varphi(t,X)).
\] 

\paragraph{Equations for the Momentum.} A direct application of the variational principle \eqref{EP_VP}--\eqref{EP_variations} which makes use of the boundary conditions for $\mathbf{u}$ and $\boldsymbol{\zeta}$ gives the equations of motion in Euler-Poincar\'e form
\begin{equation}
\label{EP-m}
\partial_t \frac{\delta\mathcal{L}}{\delta\mathbf{u}} + \pounds_{\uv} \frac{\delta\mathcal{L}}{\delta\mathbf{u}} = \sum_i \rho_i \nabla \frac{\delta\mathcal{L}}{\delta \rho_i} + s\nabla \frac{\delta\mathcal{L}}{\delta s},
\end{equation}
see \cite{HoMaRa2002}, where $\pounds_{\uv} \mathbf{m} =(\nabla \times \mathbf{m}) \times \mathbf{u} + \nabla (\uv \cdot \mathbf{m})  + \mathbf{m} \operatorname{div} \uv$ is the Lie derivative of a fluid momentum density $\mathbf{m}$ and $\frac{\delta \mathcal{L}}{\delta \mathbf{m}}, \frac{\delta \mathcal{L}}{\delta \rho_i}, \frac{\delta \mathcal{L}}{\delta s}$ are the functional derivatives of $\mathcal{L}$, see Remark \ref{remark_LD}. Introducing 
\begin{align}
\label{L-func-derivs}
\mv := \frac{\delta\mathcal{L}}{\delta \mathbf{u}},\quad\quad
B_i := -\frac{\delta\mathcal{L}}{\delta \rho_i} ,\quad\quad
T := -\frac{\delta\mathcal{L}}{\delta s}
\end{align}
equations \eqref{rho-kinematic}, \eqref{S-kinematic}, \eqref{EP-m} can be rewritten as
\begin{align}
\label{m-eqn-EPm}
&\partial _t\mv + (\nabla \times \mathbf{m}) \times \mathbf{u} + \nabla (\uv \cdot \mathbf{m})  + \mathbf{m} \operatorname{div} \uv + \sum_i \rho_i \nabla B_i + s \nabla T = 0 \\
\label{rho-eqn-EPm}
&\vspace{0.4cm}\partial _t\rho_i + \nabla \cdot (\rho_i \uv) = 0 \textcolor{white}{\sum_i}\\
\label{S-eqn-EPm}
&\partial _t s+ \nabla \cdot (s \uv) = 0.
\end{align}
As will be reviewed below, these equations are naturally connected, on the Hamiltonian side, to the Lie-Poisson formulation. 

\begin{remark}[Dual spaces and Lie derivatives]\label{remark_LD}{\rm We choose to identify the dual space to the space of vector fields tangent to the boundary, with itself, by using the duality pairing $\langle \mathbf{m},\mathbf{u}\rangle=\int_\Omega\mathbf{m}\cdot\mathbf{u}\,{\rm d}x$, where the dot is the inner product on $\mathbb{R}^3$ for $\Omega\subset\mathbb{R}^3$. If $\Omega$ is a Riemannian manifold, then the Riemannian metric must be used.

 Consistently with this choice, the functional derivative of $\mathcal{L}$ with respect to $\mathbf{u}$ is the vector field $\frac{\delta \mathcal{L}}{\delta \mathbf{u}}$ tangent to the boundary such that 
\[
\left.\frac{d}{d\varepsilon}\right|_{\varepsilon=0}\mathcal{L}[\mathbf{u}+\varepsilon\delta\mathbf{u}, \rho_i, s]=\int_\Omega \frac{\delta \mathcal{L}}{\delta \mathbf{u}}\cdot\delta\mathbf{u}\,{\rm d}x= \left\langle \frac{\delta \mathcal{L}}{\delta \mathbf{u}},\delta\mathbf{u}\right\rangle,
\]
for arbitrary vector field $\delta\mathbf{u}$ parallel to the boundary.
Such a functional derivative may or may not exist. The Lie derivative $\pounds_\mathbf{u}\mathbf{m}$ of a fluid momentum density $\mathbf{m}$ along a vector field $\mathbf{u}$ tangent to the boundary satisfies 
\[
%\left\langle\pounds_\mathbf{u}\mathbf{m},\mathbf{v}\right\rangle
\int_\Omega \pounds_\mathbf{u}\mathbf{m}\cdot\mathbf{v} {\rm d}x
= \int_\Omega \mathbf{m}\cdot ( \mathbf{v}\cdot\nabla\mathbf{u} -  \mathbf{u}\cdot\nabla\mathbf{v}){\rm d}x,\quad
%= \left\langle \mathbf{m}, \mathbf{v}\cdot\nabla\mathbf{u} -  \mathbf{u}\cdot\nabla\mathbf{v}\right\rangle,\quad 
\text{for all $\mathbf{v}$}.
\]
The other functional derivatives are defined as
\[
\left.\frac{d}{d\varepsilon}\right|_{\varepsilon=0}\mathcal{L}[\mathbf{u}, \rho_i+\varepsilon\delta\rho_i, s]=\int_\Omega \frac{\delta \mathcal{L}}{\delta \rho_i}\delta\rho_i{\rm d}x,\quad \left.\frac{d}{d\varepsilon}\right|_{\varepsilon=0}\mathcal{L}[\mathbf{u}, \rho_i, s+\varepsilon\delta s]=\int_\Omega \frac{\delta \mathcal{L}}{\delta s}\delta s{\rm d}x.
\]}\end{remark}

%\paragraph{Rotation}
%Following standard procedure, rotation is introduced into the Lagrangian by adding a term as
%\begin{equation}
%\Ll^\prime = \Ll + \left< \rho \uv, \Rv \right>
%\end{equation}
%This gives
%\begin{equation}
%\vv^\prime = \frac{1}{\rho} \dede{\Ll^\prime}{\uv} = \Rv + \dede{\Ll}{\uv}
%\end{equation}
%The momentum $\mv$ and the associated velocity $\vv$ become absolute quantities, rather than relative.

\paragraph{Specific Lagrangian.}
The specific Lagrangian that we use, which characterizes a rotating, multicomponent, multiphase fully compressible geophysical fluid with a single velocity and temperature, is
\begin{equation}
\label{specific-lagrangian}
\Ll[\uv,\rho_i,s] = \int_\Omega \rho \left( K + \uv \cdot \Rv - \Phi -  U \right) {\rm d}x,
\end{equation}
where $K = \frac{\uv \cdot \uv}{2}$ is the kinetic energy, $U(\alpha,\eta,q _i) = U(\frac{1}{\rho},\frac{s}{\rho},\frac{\rho_i}{\rho})$ is the internal energy and $\Phi$ is the geopotential. Following standard procedure, rotation has been introduced into the Lagrangian by adding the term $\int_\Omega \uv \cdot \Rv {\rm d}x$. The functional derivatives of $\Ll[\uv,\rho_i,s]$ are given by
\begin{align}
\label{specific-lagrangian-func-derivs}
\frac{\delta \mathcal{L}}{\delta \mathbf{u}} = \mv =  \rho (\uv + \Rv)= \rho \vv, \quad\quad
\frac{\delta \mathcal{L}}{\delta \rho_i} = - B_i = K + \uv \cdot \Rv - \Phi - \mu_i, \quad\quad
\frac{\delta \mathcal{L}}{\delta s} = - T,
\end{align}
with temperature $T = \pp{U}{s}$ and chemical potential $\mu_i = \pp{U}{q _i}$. For a single component fluid $\mu_i$ becomes the Gibbs free energy $G = U + p \alpha - \eta T$, with $p = -\pp{U}{\alpha}$ the pressure.

\section{Hamiltonian Formulations for Reversible Dynamics}
\label{hamiltonian-approach}

To the classical Hamilton principle \eqref{HP} in Lagrangian description, is naturally associated a Hamiltonian formulation in terms of the canonical Poisson bracket\footnote{The canonical Poisson bracket is formally given by $\{F, H\}_{\rm can}= \frac{\partial F}{\partial \varphi}\cdot\frac{\partial H}{\partial p}- \frac{\partial F}{\partial p}\cdot\frac{\partial H}{\partial \varphi}$.},
\begin{equation}\label{can_Poisson}
\dd{F}{t} = \{F,H\}_{\rm can}
\end{equation}
for the Hamiltonian $H[\varphi,p]$ defined from $L[\varphi,\dot\varphi]$ via the Legendre transform as
\begin{equation}\label{LT_Lagr}
H[\varphi, p]=\int_\Omega (p\cdot\dot\varphi ){\rm d}X - L[\varphi, \dot\varphi],
\end{equation}
where $p=\frac{\partial L}{\partial \dot\varphi}$ is the fluid momentum in the Lagrangian description and we assumed that the Lagrangian is regular. Recall that the Hamiltonian formulation \eqref{can_Poisson} gives the evolution of an arbitrary functional $F=F[\varphi, p]$.

In a similar way with the variational formulation in Section \ref{lagrangian-section}, the Hamiltonian formulation \eqref{can_Poisson} in the Lagrangian description induces a Hamiltonian formulation in the Eulerian description, given by a noncanonical Poisson bracket of Lie-Poisson type. It is sometimes advantageous to implement a change of variables from $\mv$ to $\vv$ and rewrite the Lie-Poisson bracket in curl-form. This is explored in Appendix \ref{alternative-prognostic}.

Recall that a Poisson bracket is a bilinear, antisymmetric operator on functions, that satisfies the Jacobi identity and the Leibniz rule. The noncanonical Poisson brackets for fluids gives rise to Casimir invariants, see Section \ref{reversible-casimirs}. Poisson brackets for compressible fluids, in Lie-Poisson and curl-form, were derived in \cite{MoGr1980}. The justification of the expression of Lie-Poisson brackets for fluids, as being induced by the canonical Poisson bracket in the Lagrangian description via reduction by relabelling symmetries is developed in \cite{MaRaWe1984,MaWe1983}. More details on Hamiltonian methods in geophysical fluids can be found in \cite{Shepherd1990} or other standard texts on the subject. For binary fluids, the Hamiltonian formulation using curl-form and Lie-Poisson brackets can be found in \cite{Bannon2003}, using the variable sets $(\vv,\rho,q ,\eta)$ (curl-form) or $(\mv,\rho,\rho_s,s)$ (Lie-Poisson), for $\rho_s=\rho q$. In the present paper, an extension of the Lie-Poisson brackets to multicomponent fluids  with slightly different prognostic variables $(\mv,\rho_i,s)$ is made.  A direct extension of \cite{Bannon2003} to the case of additional components is found in Appendix \ref{alt-bannon-prongostic}.

\subsection{Lie-Poisson Formulation}
\label{lie-poisson-formulation}

\paragraph{Hamiltonian function.} Given the Lagrangian $\Ll[\uv,\rho_i,s]$ of the multicomponent fluid in the Eulerian description, the Hamiltonian $\Hmh[\mv,\rho_i,s]$ (which for a rigid lid is equal to the total energy) is obtained by a Legendre transform as follows
\begin{equation}\label{Legendre_transf}
\Hmh[\mathbf{m},\rho_i,s] = \int_\Omega  \uv \cdot \frac{\delta \mathcal{L}}{\delta \mathbf{u}}{\rm d}x - \mathcal{L}[\mathbf{u},\rho_i, s] ,
\end{equation}
where $\mathbf{u}$ is such that $\frac{\delta \mathcal{L}}{\delta \mathbf{u}}=\mathbf{m}$. This is the Eulerian version of the Legendre transform \eqref{LT_Lagr}. We thus have the following relations
\begin{align}
\label{dHdx}
\frac{\delta \mathsf{H}}{\delta \mathbf{m}}  = \uv, \quad\quad
\frac{\delta \mathsf{H}}{\delta \rho_i}  =- \frac{\delta \mathcal{L}}{\delta \rho_i} =B_i,\quad\quad
 \frac{\delta \mathsf{H}}{\delta s} = - \frac{\delta \mathcal{L}}{\delta s}=T,
\end{align}
see \eqref{L-func-derivs}, where the functional derivatives of $\mathsf{H}$ are defined similarly as in Remark \ref{remark_LD}. In particular, $\frac{\delta \mathsf{H}}{\delta \mathbf{m}}\cdot\hatnv=0$. For the specific Lagrangian (\ref{specific-lagrangian}), this gives
\[
\mathsf{H}[\mathbf{m},\rho_i,s]  = \int_\Omega \rho \left[ K + \Phi + U \right] {\rm d}x,
\]
where $K$ is written in terms of $\mathbf{m}$ as $K=\frac{1}{2\rho^2}|\mathbf{m}-\rho\mathbf{R}|^2$. The functional derivatives are computed as
\begin{align}
\label{actual-func-derivs}
\frac{\delta \mathsf{H}}{\delta\mathbf{m}} = \mathbf{u} ,\quad\quad
\frac{\delta \mathsf{H}}{\delta\rho_i} = B_i = -K - \uv \cdot \Rv + \Phi + \mu_i, \quad\quad
\frac{\delta \mathsf{H}}{\delta s} = T.
\end{align}
in agreement with \eqref{specific-lagrangian-func-derivs}.

\paragraph{The Lie-Poisson Bracket.} %A opposed to the Poisson bracket in Lagrangian description, the Poisson bracket in Eulerian description is noncanonical, more precisely, it has a Lie-Poisson form.
The Eulerian version of the canonical Poisson formulation \eqref{can_Poisson} is given by
\[
\dd{\mathsf{F}}{t} = \{\mathsf{F},\mathsf{H}\},
\]
for arbitrary functionals $\Fmh[\mathbf{m},\rho_i,s]$, where $\{\,,\}$ is the noncanonical Lie-Poisson bracket
\begin{eqnarray}
\label{lie-poisson-bracket}
\{\Amh,\Bmh\} =  \{\Amh,\Bmh\}_M +\sum_i \{\Amh,\Bmh\}_{R_i} + \{\Amh,\Bmh\}_S,
\end{eqnarray}
with the three terms
\begin{align}
 \label{M-m-bracket}
 \{\Amh,\Bmh\}_M &= -\int_\Omega \mv \cdot \left( \dede{\Amh}{\mv} \cdot \nabla  \dede{\Bmh}{\mv} -  \dede{\Bmh}{\mv} \cdot \nabla \dede{\Amh}{\mv} \right) {\rm d}x \\
 \label{Ri-m-bracket}
\{\Amh,\Bmh\}_{R_i} &=  - \int_\Omega \rho_i\left(\dede{\Amh}{\mv} \cdot \nabla \frac{\delta \mathsf{B}}{\delta \rho_i} - \dede{\Bmh}{\mv} \cdot \nabla \frac{\delta \mathsf{A}}{\delta \rho_i}\right){\rm d}x \\
\label{S-m-bracket}
\{\Amh,\Bmh\}_S &= - \int_\Omega s\left(\dede{\Amh}{\mv} \cdot \nabla \frac{\delta \mathsf{B}}{\delta s} - \dede{\Bmh}{\mv} \cdot \nabla \frac{\delta \mathsf{A}}{\delta s}\right){\rm d}x.
\end{align}
In a similar way with the single component fluid, this expression of the noncanonical Poisson bracket can be directly deduced from the canonical Poisson bracket in Lagrangian description by using the process of Poisson reduction by relabelling symmetries as in \cite{MaRaWe1984,MaWe1983}.

\paragraph{Equations of Motion.}
Inserting the functional derivatives \eqref{dHdx} into the Lie-Poisson bracket \eqref{lie-poisson-bracket} and integrating by parts as needed gives the equations of motion as
\begin{align*}
&\partial_t \mv + \pounds_\mathbf{u}{\mv} +  s \nabla T + \sum_i \rho_i \nabla B_i =0 \\
&\partial_t \rho_i + \nabla \cdot (\rho _i\uv) = 0 \textcolor{white}{\sum_i}\\
&\partial_t s + \nabla \cdot (s \uv) = 0
\end{align*}
which are equivalent to \eqref{m-eqn-EPm}-\eqref{S-eqn-EPm}. By inserting the actual values for functional derivatives \eqref{actual-func-derivs}, the more common form
\begin{align*}
&\partial_t \mv +\pounds_{\uv} \mv - \rho \nabla (K + \uv \cdot \Rv) + \rho \nabla \Phi + \nabla p = 0 \\
&\partial_t \rho_i + \nabla \cdot (\rho_i \uv) = 0 \\
&\partial_t s + \nabla \cdot (s  \uv) = 0
\end{align*}
is obtained. Here we have used
\[
\sum_i \rho_i \nabla (\Phi + \mu_i) + s \nabla T = \sum_i \rho_i \nabla \Phi + \sum_i \rho_i \nabla \mu_i + s \nabla T = \rho \nabla \Phi + \nabla p
\]
since $\sum_i \rho_i = \rho$ and $\sum_i \rho_i \nabla \mu_i +s \nabla T = \nabla p$ by $\rho$ times (\ref{gibbs-duhem}). 

\subsection{Conserved Quantities and Casimirs}
\label{reversible-casimirs}

The equations of motion have at least three types of conserved quantities: the Hamiltonian $\Hmh$, the Casimirs $\Cmh$ and the linear/angular momentum. The linear and angular momenta arise from translational and rotational symmetries, respectively, via Noether's theorem, and are not discussed further. See \cite{Shepherd1990} for more details. Here we focus on the Hamiltonian $\Hmh$ and the Casimirs $\Cmh$.

\paragraph{Hamiltonian.}
By virtue of the anti-symmetry of the Poisson brackets, the equations conserve the Hamiltonian $\Hmh$, which is the total energy for a domain $\Omega$ with a rigid lid.

\paragraph{Casimirs.}
Casimirs $\Cmh$ are functionals which lie in the null space of the Poisson brackets, that is, $\{ \Amh , \Cmh \} = 0$ for any functional $\Amh$. One Casimir for the multicomponent system is
\[
\Cmh_1[\mv,\rho_i,s] = \int_\Omega \rho f(\eta,q_i) {\rm d}x,
\]
where $f$ is an arbitrary function of $\eta=\frac{s}{\rho}$ and $q_i=\frac{\rho_i}{\rho}$, $i=1,...,n$. 

\paragraph{Proof}
The functional derivatives of $\Cmh_1[\mv,\rho_i,s]$ are
\begin{align}
\label{Ch1-fd}
\frac{\delta \mathsf{C}_1}{\delta \mathbf{m}} = 0 \quad\quad
\frac{\delta \mathsf{C}_1}{\delta \rho_i} = f - \eta\partial_\eta f - \sum_{j} q_{j} \partial_{q_{j}}f +  \partial_{q_{i}}f \quad\quad
\frac{\delta \mathsf{C}_1}{\delta s} = \partial_\eta f .
\end{align}
Casimirs must satisfy  $\{ \Amh , \Cmh \} = 0$, which gives
\[
\int_\Omega \dede{\Amh}{\mv} \cdot \left[ \sum_i \rho_i \nabla\frac{\delta \mathsf{C}_1}{\delta \rho_i} + s \nabla\frac{\delta \mathsf{C}_1}{\delta s} \right] {\rm d}x = 0.
\]
Since this must hold for arbitrary $\Amh$, this implies that
\begin{equation}
\label{Ch1-eqn}
\sum_i \rho_i \nabla\frac{\delta \mathsf{C}_1}{\delta \rho_i} + s \nabla\frac{\delta \mathsf{C}_1}{\delta s} =0.
\end{equation}
Straightforward calculation with \eqref{Ch1-fd} and use of the chain rule verifies that \eqref{Ch1-eqn} holds.

This Casimir is a straightforward generalization of the Casimir $\Ch_1$ from \cite{Bannon2003} to the case of $n$ components and slightly different prognostic variables, and it is a consequence of material conservation of entropy $\eta$ and concentration $q_i$. Important special cases are total mass of component $i$ for $f=q_i$, total mass for $f=1$, and total entropy for $f=\eta$. 

\paragraph{Total Entropy.} Since it plays a prominent role in the formulation of the single and double generator dissipation brackets, we will denote the total entropy Casimir ($\Cmh_1$ with $f=\eta$) as $\Smh[s] = \int \rho \eta {\rm d}x= \int s{\rm d}x$, which has functional derivatives
\[
\frac{\delta\mathsf{S}}{\delta s} = 1, \qquad \frac{\delta\mathsf{S}}{\delta \rho_i} =0, \qquad  \frac{\delta\mathsf{S}}{\delta \mathbf{m}} = 0.
\]

\paragraph{Potential Vorticity Casimir.}
Unlike the single component case, $\Cmh = \int_\Omega \rho f(\eta,q){\rm d}x$ is not a Casimir, where $q= \frac{\nabla\eta \cdot \operatorname{curl}\mathbf{v}}{\rho}$ is the potential vorticity. This is because $q$ is no longer materially conserved (see Appendix \ref{pv-kelvin-circulation}). No claim is made that this is an exhaustive set of Casimirs for the multicomponent system. For a binary system, there are at least two additional Casimirs, discussed further in \cite{Bannon2003}.

\section{Single and Double Generator Bracket Formulations}
\label{bracket-formulations}

\paragraph{Variational formulation for nonequilibrium thermodynamics.} A variational formulation for systems with irreversible processes was developed in \cite{GBYo2017a,GBYo2017b}, and applied to moist, multicomponent geophysical fluids in \cite{GayBalmaz2017}. This variational formulation extends the Hamilton principle \eqref{HP} to include irreversible processes, using a systematic structure that is common to finite dimensional and continuum thermodynamic systems. It relies on the specification of entropy generation rates interpreted as a constraint in the variational principle, and on the introduction of the associated concept of \textit{thermodynamic displacement}, as we will review below. In a similar way with the Hamilton principle, this variational formulation also has an Eulerian version that extends the Euler-Poincar\'e approach to irreversible processes.

In this section we will use this variational formulation to systematically develop bracket formulations that incorporate irreversible processes. These formulations are composed of a Poisson bracket for the reversible dynamics, and a \textit{dissipation} bracket for the irreversible dynamics.

\paragraph{Bracket Formalism.}

The inclusion of dissipative or irreversible phenomena in Hamilton's equations through a modification of the Poisson bracket has been initiated by \cite{Gr1984,Ka1984,Mo1984a,Mo1984b}.
There are two main approaches to the dissipation bracket in the literature, depending on which \textit{generating function} they use (see below): the \textit{single generator} and \textit{double generator} formulations.  Depending on the type of system being simulated, the relevant entropy can be defined such that the inequalities below are $\leq$ rather than $\geq$. 

\medskip

In the single generator formalism, \cite{BeEd1994,EdBe1991b,EdBe1991a}, the evolution of an arbitrary functional $\Fmh$ is governed by 
\[
\dd{\Fmh}{t} = \{\Fmh,\Hmh\} + [\Fmh,\Hmh],
\]
where the dissipation bracket $[\Fmh,\Hmh]$ is linear in $\mathsf{F}$ and a derivation in $\Fmh$, can be nonlinear in $\Hmh$, and satisfies $[\Hmh,\Hmh] = 0$ and $[\Smh,\Hmh] \geq 0$. These last two requirements are the first and second laws of thermodynamics, respectively. Since both the reversible (Poisson) and dissipation brackets use the same generator $\Hmh$, this is referred to as the \textit{single generator formalism}. 

\medskip

In the double generator formalism, the evolution of an arbitrary functional $\Fmh$ is governed by 
\[
\dd{\Fmh}{t} = \{\Fmh,\Hmh\} + (\Fmh,\Smh)
\]
where the function $\mathsf{S}$ is such that $\{\mathsf{H},\mathsf{S}\}=0$, and the dissipation bracket $(\Fmh,\Smh)$ is symmetric, bilinear and satisfies the Leibniz rule, $(\Hmh,\Smh) = 0 $ and $(\Smh,\Smh) \geq 0$. These are precisely the axioms given in  \cite{Ka1984}. Since the Poisson and dissipation brackets use different generators ($\Hmh$ for Poisson and $\Smh$ for dissipation), this is referred to as the \textit{double generator formalism}. Sometimes, the stronger requirements that $\{\mathsf{A}, \mathsf{S}\}=0$, $(\Hmh,\Amh) = 0$, $(\mathsf{A},\mathsf{A})\geq 0$ for an arbitrary $\Amh$ is imposed, in which case the complete system is termed \textit{metriplectic}, \cite{Mo1986}. For example, this is what is used in the GENERIC formalism \cite{GrOtt1997,OtGr1997}. When considering macroscopic systems, typically only bilinearity, $(\Hmh,\Smh) = 0 $ and $(\Smh,\Smh) \geq 0$ seem to be required on physical grounds. A discussion of these issues, and a comparison between the single and double generator formalisms for macroscopic single component fluids and microscopic systems can be found in \cite{Ed1998,EdBeOt1998}.

\subsection{Variational Formulation with Irreversible Processes}
\label{variational-irr-section}

The variational formulation of nonequilibrium thermodynamics developed in \cite{GBYo2017a,GBYo2017b} is an extension of the Hamilton principle \eqref{HP} that includes the irreversible processes. This is done by imposing two constraints on the variational principle, a constraint on the critical curve (the phenomenological constraint) and a constraint on the variations (the variational constraint). As we will see below, the relation between these two constraints and the expression of the constraint follow a very systematic construction, that turns out to be common to finite dimensional and continuum thermodynamic systems. It is based on the concept of \textit{thermodynamic displacement} of an irreversible process, defined such that its time derivative is the affinity of the process. Formally, if $J_\alpha$, $X^\alpha$ are the thermodynamic flux and the thermodynamic affinity of the process $\alpha$, then the thermodynamic displacement is $\Lambda^\alpha$ such that $\dot\Lambda^\alpha=X^\alpha$. The internal entropy production is $-\frac{1}{T}\sum_\alpha J_\alpha X^\alpha=-\frac{1}{T} \sum_\alpha J_\alpha \dot\Lambda^\alpha$. The phenomenological constraint and variational constraints are related as
\[
J_\alpha \dot\Lambda^\alpha \leadsto J_\alpha \delta\Lambda^\alpha,
\]
for adiabatically closed systems, see \cite{GBYo2018} for open systems.

We now recall the variational formulation directly in the Eulerian description and refer to \cite{GayBalmaz2017,GBYo2017b} for the Lagrangian description. In our case, the thermodynamic displacements are the thermal displacement $\gamma(t,x)$ and matter displacements $w_i(t,x)$.
The thermodynamic fluxes are the viscous stress tensor $\sigmafr$, the diffusion flux $\mathbf{j}_i$ for component $i$, the conversion rate $j_i$ for component $i$, and the entropy flux $\mathbf{j}_s$. The domain is assumed to be adiabatically closed, and therefore $\mathbf{j}_s \cdot \hatnv = \mathbf{j}_i \cdot \hatnv = 0$ on $\partial \Omega$. Also, we assume $\mathbf{u} =0$ on $\partial \Omega$. This is a stronger condition than reversible dynamics, which requires only $\mathbf{u} \cdot \hatnv = 0$ on $\partial \Omega$. This distinction between reversible and irreversible boundary conditions occurs also in the Navier-Stokes equations for incompressible flow. The diffusion fluxes $\mathbf{j}_i$ and conversion rates $j_i$ are subject to the mass control conditions $\sum_i \mathbf{j}_i= 0$ and $\sum_i j_i = 0$. 

The variational formulation reads
\begin{equation}\label{Review_GBYo} 
\delta   \int_0^T \Big[\mathcal{L}[\mathbf{u},\rho_i, s] +\int_\Omega \sum_i \rho _i D_t w_i {\rm d}x+  \int_\Omega(s- \sigma ) D_t \gamma    {\rm d}x \Big] {\rm d}t =0,
\end{equation}
subject to the \textit{phenomenological constraint}
\begin{equation}\label{KC_GBYo}
\frac{\partial \mathcal{L} }{\partial s}\bar D_t \sigma= -\boldsymbol{\sigma}^{\rm fr}: \nabla\mathbf{u}  + \mathbf{j} _s \cdot \nabla D_t\gamma  + \sum_i( \mathbf{j} _i \cdot \nabla D_tw _i+  j_i D_t w _i )
\end{equation}
and with respect to variations subject to $ \delta \mathbf{u} =\partial _t \boldsymbol{\zeta} + \mathbf{u} \cdot \nabla \boldsymbol{\zeta} - \boldsymbol{\zeta}\cdot \nabla \mathbf{u} $ and to the \textit{variational constraint}
\begin{equation}\label{VC_GBYo}
\frac{\partial \mathcal{L} }{\partial s}\bar D_\delta \sigma= -\boldsymbol{\sigma}^{\rm fr}: \nabla\boldsymbol{\zeta}  + \mathbf{j} _s \cdot \nabla D_\delta\gamma  + \sum_i( \mathbf{j} _i \cdot \nabla D_\delta w _i+  j_i D_\delta w _i )
\end{equation} 
with $ \delta w _i $, $ \delta \gamma $, and $\boldsymbol{\zeta} $ vanishing at $t=0,T$.

We used the Lagrangian derivatives and variations $D_tf:= \partial _t f+ \mathbf{u}\cdot \nabla f$, $\bar D_tf:= \partial _t f+ \nabla \cdot(f\mathbf{u})$, $D_\delta f:= \delta f+ \boldsymbol{\zeta}\cdot \nabla f$, $\bar D_ \delta f:=\delta f+ \nabla \cdot(f \boldsymbol{\zeta})$. One passes from the phenomenological constraint \eqref{KC_GBYo} to the variational constraint \eqref{VC_GBYo} by replacing time derivatives by delta variations. In absence of irreversible processes, both constraints disappear and the variational formulation reduces to the Euler-Poincar\'e formulation. A direct application of \eqref{Review_GBYo}--\eqref{VC_GBYo} yields the system
\begin{equation}\label{system_Eulerian_atmosphere_moist_L} 
\left\{
\begin{array}{l}
\vspace{0.2cm}\displaystyle
\partial _t\frac{\delta \mathcal{L} }{\delta \mathbf{u} } + \pounds _ \mathbf{u}  \frac{\delta \mathcal{L} }{\delta \mathbf{u} } = \sum_i\rho _i \nabla \frac{\delta \mathcal{L} }{\delta \rho _i }+ s \nabla \frac{\delta \mathcal{L} }{\delta s }+ \nabla\cdot  \boldsymbol{\sigma}  ^{\rm fr}\\
\vspace{0.2cm}\displaystyle\frac{\delta \mathcal{L} }{\delta s } (\bar D_t s + \nabla\cdot \mathbf{j} _s ) = - \boldsymbol{\sigma} ^{\rm fr} \!: \!\nabla \mathbf{v} - \mathbf{j} _s\! \cdot  \!\nabla \frac{\delta \mathcal{L} }{\partial s }- \sum_i\left(\mathbf{j} _i \!\cdot \! \nabla \frac{\delta \mathcal{L} }{\delta \rho _i }+j _i  \frac{\delta \mathcal{L} }{\delta \rho_i} \right) \\
\displaystyle \bar D_ t\rho _i + \nabla\cdot \mathbf{j} _i=j _i,
\end{array}
\right.
\end{equation}
see \cite{GayBalmaz2017,GBYo2017b} for detailed computations.
These are the general equations for a fluid with Lagrangian $\mathcal{L}[\mathbf{u},\rho_i,s]$ subject to the irreversible processes of viscosity, heat conduction, diffusion, and phase changes. They clearly recover \eqref{EP-m} in absence of the irreversible processes. The system is closed by specifying a relationship, or \textit{parameterizing}, the thermodynamic fluxes ($\boldsymbol{\sigma}^{\rm fr}$, $\mathbf{j}_i$, $j_i$, $\mathbf{j}_s$) in terms of the thermodynamic forces ($\operatorname{Def}\mathbf{u}= \frac{1}{2}(\nabla\mathbf{u}+\nabla\mathbf{u}^\mathsf{T})$, $\nabla T$, $\nabla \mu_i$, $\mu_i$), see \cite{GayBalmaz2017}. More details on this can be found in Section \ref{parameterizing-fluxes}, where we present two different approaches. For the Lagrangian \eqref{specific-lagrangian} of the rotating compressible multicomponent fluid, \eqref{system_Eulerian_atmosphere_moist_L} yields
\begin{equation}\label{system_Eulerian_atmosphere_moist_L_final} 
\left\{
\begin{array}{l}
\vspace{0.2cm}\displaystyle
\rho(\partial _t \mathbf{u} + \mathbf{u} \cdot\nabla\mathbf{u} + 2 \boldsymbol{\Omega}\times \mathbf{u}) = - \nabla p - \rho \nabla \Phi+ \nabla\cdot  \boldsymbol{\sigma}  ^{\rm fr}\\
\displaystyle T (\bar D_t s + \nabla\cdot \mathbf{j} _s ) =  \boldsymbol{\sigma} ^{\rm fr} \!: \!\nabla \mathbf{u} - \mathbf{j} _s\! \cdot  \!\nabla T -  \sum_i\left(\mathbf{j} _i \!\cdot \! \nabla \mu _i +j _i  \mu_i \right) \\
\displaystyle \bar D_ t\rho _i + \nabla\cdot \mathbf{j} _i=j _i.
\end{array}
\right.
\end{equation}

Let us consider the Hamiltonian $\mathsf{H}[\mathbf{m},\rho_i, s]$ associated to $\mathcal{L}[\mathbf{u},\rho_i, s]$ via the Legendre transform as in \eqref{Legendre_transf}. Using \eqref{dHdx} we can directly rewrite the equations \eqref{system_Eulerian_atmosphere_moist_L} in terms of $\mathsf{H}$ and the variables $(\mathbf{m},\rho_i, s)$ as
\begin{equation}\label{system_Eulerian_atmosphere_moist_H} 
\left\{
\begin{array}{l}
\vspace{0.2cm}\displaystyle
\partial _t\mathbf{m} + \pounds _ { \frac{\delta \mathsf{H} }{\delta \mathbf{m} }}  \mathbf{m} = - \sum_i\rho _i \nabla \frac{\delta\mathsf{H}}{\delta \rho _i } - s \nabla \frac{\delta \mathsf{H} }{\delta s }+ \nabla\cdot  \boldsymbol{\sigma}  ^{\rm fr}\\
\vspace{0.2cm}\displaystyle\frac{\delta \mathsf{H} }{\delta s } \left(\partial_t s +\nabla\cdot \left( s \frac{\delta \mathsf{H} }{\delta \mathbf{m} }\right)+ \nabla\cdot \mathbf{j} _s \right) = \boldsymbol{\sigma} ^{\rm fr} \!: \!\nabla  \frac{\delta \mathsf{H} }{\delta \mathbf{m} } - \mathbf{j} _s\! \cdot  \!\nabla  \frac{\delta \mathsf{H} }{\delta s }- \sum_i\left(\mathbf{j} _i \!\cdot \! \nabla  \frac{\delta \mathsf{H} }{\delta \rho _i }+j _i   \frac{\delta \mathsf{H} }{\delta\rho_i} \right) \\
\displaystyle \partial_t\rho _i + \nabla\cdot \left( \rho_i \frac{\delta \mathsf{H} }{\delta \mathbf{m} }\right)+ \nabla\cdot \mathbf{j} _i=j _i.
\end{array}
\right.
\end{equation} 
From this, the evolution of an arbitrary functional $\Amh[\mv,\rho_i,s]$ is
\begin{equation}\label{A-evolution}
\begin{aligned}
\frac{d}{dt} \mathsf{A}  = &\{\mathsf{A}, \mathsf{H}\} + \int_\Omega \frac{\frac{\delta \mathsf{A}}{\delta s}}{\frac{\delta \mathsf{H}}{\delta s}}\Big(\boldsymbol{\sigma} ^{\rm fr} \!: \!\nabla  \frac{\delta \mathsf{H} }{\delta \mathbf{m} } - \mathbf{j} _s\! \cdot  \!\nabla  \frac{\delta \mathsf{H} }{\delta s }- \sum_i\mathbf{j} _i \!\cdot \! \nabla  \frac{\delta \mathsf{H} }{\delta \rho _i } - \sum_i j _i   \frac{\delta \mathsf{H} }{\delta\rho_i} \Big){\rm d}x\\
& - \int _\Omega\nabla \frac{\delta \mathsf{A}}{\delta\mathbf{m}} : \sigmafr {\rm d}x + \int _\Omega\nabla \frac{\delta \mathsf{A}}{\delta s} \cdot \mathbf{j}_s{\rm d} x +\sum_i \int _\Omega\nabla \frac{\delta \mathsf{A}}{\delta \rho_i} \cdot \mathbf{j}_i{\rm d} x+\sum_i \int _\Omega \frac{\delta \mathsf{A}}{\delta \rho_i} j_i{\rm d} x.
\end{aligned}
\end{equation}
Below, we will show how the variational formalism directly gives rise to single and double generator dissipation brackets.

\subsection{Single Generator Bracket}
\label{single-gen-bracket-section}
Now assume that the thermodynamic fluxes can be parameterized in terms of the forces in an arbitrary way, and note that the thermodynamic forces  ($\operatorname{Def}\mathbf{u}= \frac{1}{2}(\nabla\mathbf{u}+\nabla\mathbf{u}^\mathsf{T})$, $\nabla T$, $\nabla \mu_i$, $\mu_i$) are in fact functions of $\frac{\delta\mathsf{H}}{\delta \mathbf{m}}$, $\frac{\delta\mathsf{H}}{\delta \rho_i}$, $\frac{\delta\mathsf{H}}{\delta s}$. Writing $\frac{\delta\mathsf{H}}{\delta x}$ the set of all these derivatives, we can write $\sigmafr = \sigmafr(\frac{\delta\mathsf{H}}{\delta x})$, $\mathbf{j}_s = \mathbf{j}_s(\frac{\delta\mathsf{H}}{\delta x})$, $\mathbf{j}_i = \mathbf{j}_i(\frac{\delta\mathsf{H}}{\delta x})$ and $j_i = j_i(\frac{\delta\mathsf{H}}{\delta x})$. Then from \eqref{A-evolution} we obtain directly the single generator bracket
\begin{equation}\label{single-generator-bracket}
\begin{aligned}
{[}\mathsf{A},\mathsf{H}{]}=& - \int _\Omega\nabla \frac{\delta \mathsf{A}}{\delta\mathbf{m}} : \sigmafr  \Big(\frac{\delta \mathsf{H}}{\delta x}\Big){\rm d}x\\
&+\int_\Omega \frac{\frac{\delta \mathsf{A}}{\delta s}}{\frac{\delta \mathsf{H}}{\delta s}}\left(\boldsymbol{\sigma} ^{\rm fr} \Big(\frac{\delta \mathsf{H}}{\delta x}\Big)\!: \!\nabla  \frac{\delta \mathsf{H} }{\delta \mathbf{m} } - \mathbf{j} _s \Big(\frac{\delta \mathsf{H}}{\delta x}\Big)\! \cdot  \!\nabla  \frac{\delta \mathsf{H} }{\delta s } \textcolor{white}{\sum_i}\right.\\
&\hspace{2cm}\left.- \sum_i\mathbf{j} _i \Big(\frac{\delta \mathsf{H}}{\delta x}\Big) \!\cdot \! \nabla  \frac{\delta \mathsf{H} }{\delta \rho _i } - \sum_i j _i  \Big(\frac{\delta \mathsf{H}}{\delta x}\Big)  \frac{\delta \mathsf{H} }{\delta\rho_i} \right){\rm d}x\\
&+ \int _\Omega\nabla \frac{\delta \mathsf{A}}{\delta s} \cdot \mathbf{j}_s \Big(\frac{\delta \mathsf{H}}{\delta x}\Big){\rm d} x +\sum_i \int _\Omega\nabla \frac{\delta \mathsf{A}}{\delta \rho_i} \cdot \mathbf{j}_i \Big(\frac{\delta \mathsf{H}}{\delta x}\Big){\rm d} x+\sum_i \int _\Omega \frac{\delta \mathsf{A}}{\delta \rho_i} j_i \Big(\frac{\delta \mathsf{H}}{\delta x}\Big){\rm d} x.
\end{aligned}
\end{equation}
This bracket is linear in $\Amh$, nonlinear in $\Hmh$, and satisfies $[\Hmh,\Hmh] =0$ and $[\Smh,\Hmh]\geq 0$. The proof of the first three is left to the interested reader, noting that $[\Hmh,\Hmh] =0$ relies only cancellation of terms and is independent of the parameterization. The last condition, $[\Smh,\Hmh]\geq 0$, requires that
\[
\int_\Omega \frac{1}{\frac{\delta \mathsf{H}}{\delta s}}\left(\boldsymbol{\sigma} ^{\rm fr} \Big(\frac{\delta \mathsf{H}}{\delta x}\Big)\!: \!\nabla  \frac{\delta \mathsf{H} }{\delta \mathbf{m} } - \mathbf{j} _s \Big(\frac{\delta \mathsf{H}}{\delta x}\Big)\! \cdot  \!\nabla  \frac{\delta \mathsf{H} }{\delta s }- \sum_i\mathbf{j} _i \Big(\frac{\delta \mathsf{H}}{\delta x}\Big) \!\cdot \! \nabla  \frac{\delta \mathsf{H} }{\delta \rho _i } - \sum_i j _i  \Big(\frac{\delta \mathsf{H}}{\delta x}\Big)  \frac{\delta \mathsf{H} }{\delta\rho_i} \right){\rm d}x\geq 0,
\]
which will be satisfied only for certain parameterizations. See Section \ref{parameterizing-fluxes} for more details. In the case of a single component all of the terms involving $\mathbf{j}_i$ and $j_i$ drop out, and \eqref{single-generator-bracket} yields the single generator bracket from \cite[(7.2-1)]{BeEd1994} and \cite[(17)]{Ed1998}, where they have assumed relationships of the form $\sigmafr (\frac{\delta \mathsf{H}}{\delta x}) = \mathsf{Q} \cdot \nabla \frac{\delta \mathsf{H}}{\delta \mathbf{m}}$ and $\mathbf{j}_s(\frac{\delta \mathsf{H}}{\delta x}) = \alpha \cdot \nabla \frac{\delta \mathsf{H}}{\delta s}$, for some tensors $\mathsf{Q}$ and $\alpha$. In the multicomponent case, in absence of the conversion rates $j_i$, \eqref{single-generator-bracket} recovers the single generator bracket from \cite[(7.3-7)]{BeEd1994}.

\subsection{Double Generator Bracket}\label{subsec_double}
\label{double-gen-bracket-section}

Now consider \eqref{A-evolution} as being the result of $(\Amh,\Smh)$ for a double generator bracket using the total entropy $\Smh$, with $\Hmh$ not treated as argument to the bracket. Replacing $\Smh$ by an arbitrary functional $\Bmh$ and symmetrizing gives
\begin{equation}
\label{double-generator-bracket}
\begin{aligned}
(\mathsf{A},\mathsf{B})=& - \int_\Omega \nabla \frac{\delta \mathsf{A}}{\delta \mathbf{m}}: \sigmafr \frac{\delta \mathsf{B}}{\delta s} {\rm d}x - \int _\Omega\nabla \frac{\delta \mathsf{B}}{\delta \mathbf{m}}: \sigmafr \frac{\delta \mathsf{A}}{\delta s}{\rm d}x \\
&+ \int_\Omega \frac{1}{\frac{\delta \mathsf{H}}{\delta s}} \left(\sigmafr : \nabla \frac{\delta \mathsf{H}}{\delta \mathbf{m}} - \mathbf{j}_s\cdot \nabla \frac{\delta \mathsf{H}}{\delta s}- \sum_i \mathbf{j}_i \cdot \nabla \frac{\delta \mathsf{H}}{\delta \rho_i} - \sum_i  j_i \frac{\delta \mathsf{H}}{\delta \rho_i} \right) \frac{\delta \mathsf{A}}{\delta s} \frac{\delta \mathsf{B}}{\delta s} {\rm d}x\\
&+ \int_\Omega \nabla \frac{\delta \mathsf{A}}{\delta s} \cdot \mathbf{j}_s \frac{\delta \mathsf{B}}{\delta s}{\rm d}x + \int_\Omega \nabla  \frac{\delta \mathsf{B}}{\delta s}\cdot \mathbf{j}_s \frac{\delta \mathsf{A}}{\delta s} {\rm d}x\\
&+ \sum_i \int_\Omega \nabla \frac{\delta \mathsf{A}}{\delta \rho_i} \cdot \mathbf{j}_i \frac{\delta \mathsf{B}}{\delta s}{\rm d}x + \sum_i \int_\Omega \nabla \frac{\delta \mathsf{B}}{\delta\rho_i} \cdot \mathbf{j}_i \frac{\delta \mathsf{A}}{\delta s}{\rm d}x \\
&+  \sum_i \int _\Omega \frac{\delta \mathsf{A}}{\delta \rho_i} j_i \frac{\delta \mathsf{B}}{\delta s}{\rm d}x+\sum_i \int_\Omega \frac{\delta \mathsf{B}}{\delta \rho_i} j_i \frac{\delta \mathsf{A}}{\delta s}{\rm d}x.
\end{aligned}
\end{equation}
This bracket is bilinear, symmetric and satisfies $(\Hmh,\Smh)=0$ and $(\Smh,\Smh)\geq 0$; and depends parametrically on $\Hmh$. Again, the proof of the first three properties is left to the interested reader. These are precisely the axioms given in \cite{Ka1984}. The condition for $(\Smh,\Smh) \geq 0$ is
\[
\int_\Omega \frac{1}{\frac{\delta \mathsf{H}}{\delta s}}\left(\boldsymbol{\sigma} ^{\rm fr}\!: \!\nabla  \frac{\delta \mathsf{H} }{\delta \mathbf{m} } - \mathbf{j} _s \! \cdot  \!\nabla  \frac{\delta \mathsf{H} }{\delta s }- \sum_i\mathbf{j} _i  \!\cdot \! \nabla  \frac{\delta \mathsf{H} }{\delta \rho _i } - \sum_i j _i  \frac{\delta \mathsf{H} }{\delta\rho_i} \right){\rm d}x\geq 0,
\]
which (as expected) is the same as in the single generator formulation. 
A direct check shows that $(\mathsf{A},\mathsf{S})$, for arbitrary functional $\mathsf{A}$ gives all the terms in \eqref{A-evolution}.

This bracket is not metriplectic with $(\Hmh,\Amh)=0$ for arbitrary $\Amh$, instead it gives
\begin{equation}\label{AH_term}
(\Hmh,\Amh) = - \int _\Omega \frac{\delta \mathsf{H}}{\delta s} \left(\sigmafr : \nabla \frac{\delta \mathsf{A}}{\delta \mathbf{m}} - \mathbf{j}_s\cdot \nabla \frac{\delta \mathsf{A}}{\delta s} - \sum_i \mathbf{j}_i \cdot \nabla \frac{\delta \mathsf{A}}{\delta \rho_i} - \sum_i  j_i \frac{\delta \mathsf{A}}{\delta \rho_i} \right){\rm d}x.
\end{equation}
%\begin{align}
%(\Amh,\Smh) &= \int - \nabla \dede{\Amh}{\mv} : \sigmafr + \nabla \dede{\Amh}{s} \cdot \js + \sum_i \nabla \dede{\Amh}{\rho_i} \cdot \ji + \dede{\Amh}{\rho_i} j_i \\
%&+ \frac{\dede{\Amh}{s}}{\dede{\Hmh}{s}} \left( \int \nabla \dede{\Amh}{\mv} : \sigmafr - \nabla \dede{\Amh}{s} \cdot \js - \sum_i \nabla \dede{\Amh}{\rho_i} \cdot \ji - \dede{\Amh}{\rho_i} j_i \right)
%\end{align}
In the case of a single component all of the terms involving $\mathbf{j}_i$ and $j_i$ drop out, and \eqref{double-generator-bracket} resembles a symmetrized version of equation (45) from \cite{Ed1998}; where they have assumed relationships of the form $\sigmafr (\dede{\Hmh}{x}) = \mathsf{D} \cdot \nabla \dede{\Hmh}{\mv}$ and $\js(\dede{\Hmh}{x}) = \alpha \cdot \nabla \dede{\Hmh}{s}$, for some tensors $\mathsf{D}$ and $\alpha$.

\paragraph{Metriplectic Bracket.} We shall now see that the expression \eqref{A-evolution} found via the variational formulation can also yield a double generator bracket which is metriplectic \cite{Mo1986}. As we have recalled above, in the metriplectic formalism, the symmetric bracket not only has to satisfy the conditions $(\Hmh,\Smh) = 0$ and $(\mathsf{S},\mathsf{S})\geq 0$, but also the stronger conditions  $(\Hmh,\Amh) = 0$, $(\mathsf{A},\mathsf{A})\geq 0$, for arbitrary functionals $\mathsf{A}$. These are also the conditions imposed in the GENERIC formalism, \cite{GrOtt1997,Mo1986,OtGr1997}.

In order to obtain the metriplectic bracket we consider \eqref{AH_term} and assume the following linear relations
\[
\boldsymbol{\sigma}^{\rm fr} = Q \cdot \nabla \frac{\delta h}{\delta \mathbf{m}},\qquad -\mathbf{j}_k= \sum_{l=0}^n L_{kl} \nabla \frac{\delta h}{\delta x_l},\qquad - j_i= \sum_{j=1}^n l_{ij} \frac{\delta h}{\delta\rho_j},
\]
where $L_{kl}=L_{lk}$, $l_{ij}=l_{ji}$ and $Q$ are positive semi-definite and we used the notations $x_{k=0}=s$, $x_{k=j}=\rho_j$ and $\mathbf{j}_{k=0}=\mathbf{j}_s$, $\mathbf{j}_{k=i}=\mathbf{j}_i$. This is in fact an assumption of a parameterization for the thermodynamic fluxes in terms of the thermodynamic forces. With these relations, \eqref{AH_term} becomes
\begin{equation}\label{AH_term_new}
- \int _\Omega T\Big(Q\cdot \nabla \frac{\delta \mathsf{H}}{\delta \mathbf{m}}  : \nabla \frac{\delta \mathsf{A}}{\delta \mathbf{m}} +\sum_{k,l=0}^n  L_{kl}\nabla \frac{\delta \mathsf{H}}{\delta x_k} \nabla \frac{\delta \mathsf{A}}{\delta x_l} + \sum_{i,j=1}^n l_{ij} \frac{\delta \mathsf{H}}{\delta \rho_i} \frac{\delta \mathsf{A}}{\delta \rho_j} \Big){\rm d}x=:(\!(\mathsf{A},\mathsf{H})\!).
\end{equation}
This expression is symmetric. We can thus define the symmetric bracket
\begin{equation}\label{def_mp}
(\mathsf{A},\mathsf{B})_{\rm mp}:= (\mathsf{A},\mathsf{B})-(\!(\mathsf{A},\mathsf{B})\!).
\end{equation}
We have $(\mathsf{A},\mathsf{H})_{\rm mp}= (\mathsf{A},\mathsf{H})-(\!(\mathsf{A},\mathsf{H})\!)=(\mathsf{A},\mathsf{H}) - (\mathsf{A},\mathsf{H})=0$, for all functionals $\mathsf{A}$. It remains to show that $(\mathsf{A},\mathsf{A})_{\rm mp}\geq 0$ for arbitrary functionals $\mathsf{A}$.

A long computation shows that
\begin{align*}
(\mathsf{A},\mathsf{B})_{\rm mp}=&\int _\Omega  T \left (\nabla \frac{\delta \mathsf{A}}{\delta\mathbf{m}} - \frac{1}{T} \nabla \frac{\delta \mathsf{H}}{\delta\mathbf{m}}  \frac{\delta \mathsf{A}}{\delta s} \right) :Q: \left( \nabla \frac{\delta \mathsf{B}}{\delta\mathbf{m}} - \frac{1}{T} \nabla \frac{\delta \mathsf{H}}{\delta\mathbf{m}}  \frac{\delta \mathsf{B}}{\delta s}\right){\rm d}x\\
&+  \sum_{k,l=0}^n \int _\Omega T \left(   \nabla \frac{\delta \mathsf{A}}{\delta x_k} -\frac{1}{T}  \nabla \frac{\delta \mathsf{H}}{\delta x_k} \frac{\delta \mathsf{A}}{\delta s}\right) L_{kl} \left( \nabla \frac{\delta \mathsf{B}}{\delta x_l} -\frac{1}{T}  \nabla \frac{\delta \mathsf{H}}{\delta x_l} \frac{\delta \mathsf{B}}{\delta s}\right){\rm d}x\\
&+ \sum_{i,j=1}^n \int _\Omega T \left( \frac{\delta \mathsf{A}}{\delta\rho_i} -\frac{1}{T}  \frac{\delta \mathsf{H}}{\delta\rho_i}  \frac{\delta \mathsf{A}}{\delta s} \right) l_{ij}\left(  \frac{\delta \mathsf{B}}{\delta\rho_j}-\frac{1}{T}  \frac{\delta \mathsf{H}}{\delta\rho_j}  \frac{\delta \mathsf{B}}{\delta s}\right){\rm d}x
\end{align*}
hence it follows that $(\mathsf{A},\mathsf{A})_{\rm mp}\geq 0$ for all functionals $\mathsf{A}$. This shows that
\[
\{\mathsf{A},\mathsf{B}\}+ (\mathsf{A},\mathsf{B})_{\rm mp}
\]
is a metriplectic (or GENERIC) bracket.

In the case of a single component all of the terms involving $\mathbf{j}_i$ and $j_i$ drop out and we get a metriplectic (or GENERIC) bracket for the compressible heat conducting viscous fluid. However, in this case, a simpler metriplectic bracket can be derived. Using again the expression \eqref{A-evolution} found via the variational formulation and assuming
\[
\boldsymbol{\sigma}^{\rm fr}= Q\cdot \nabla \frac{\delta \mathsf{H}}{\delta\mathbf{m}}\quad \text{and}\quad T\mathbf{j}_s = -\kappa \nabla T,
\]
the irreversible part of \eqref{A-evolution} becomes
\[
- \int_\Omega \nabla \frac{\delta \mathsf{A}}{\delta\mathbf{m}}:\boldsymbol{\sigma}^{\rm fr}{\rm d}x + \int_\Omega \frac{1}{T}\boldsymbol{\sigma}^{\rm fr}:\nabla \frac{\delta \mathsf{H}}{\delta\mathbf{m}}  \frac{\delta \mathsf{A}}{\delta s}{\rm d}x +\kappa \int_\Omega T^2\nabla\left( \frac{1}{T}\frac{\delta \mathsf{A}}{\delta s}\right)\cdot  \nabla\left( \frac{1}{T} \right){\rm d}x.
\]
From this we deduce a double generator structure $(\mathsf{A},\mathsf{B})$ by following the same steps as in the beginning of \S\ref{subsec_double}. Then, computing $(\mathsf{A},\mathsf{H})$ as in \eqref{AH_term} and proceeding as in \eqref{AH_term_new} and \eqref{def_mp}, we obtain the metriplectic structure
\begin{align*}
(\mathsf{A},\mathsf{B})_{\rm mp} = &- \int _\Omega \nabla \frac{\delta \mathsf{A}}{\delta\mathbf{m}}:\boldsymbol{\sigma}^{\rm fr} \frac{\delta \mathsf{B}}{\delta s}{\rm d}x - \int _\Omega\nabla \frac{\delta \mathsf{B}}{\delta\mathbf{m}}:\boldsymbol{\sigma}^{\rm fr}\frac{\delta \mathsf{A}}{\delta s}{\rm d}x+ \int _\Omega\frac{1}{T}\boldsymbol{\sigma}^{\rm fr}:\nabla \frac{\delta \mathsf{H}}{\delta\mathbf{m}}  \frac{\delta \mathsf{A}}{\delta s}\frac{\delta \mathsf{B}}{\delta s}{\rm d}x\\
&+\kappa \int _\Omega T^2\nabla\left( \frac{1}{T}\frac{\delta \mathsf{A}}{\delta s}\right)\cdot  \nabla\left( \frac{1}{T}\frac{\delta \mathsf{B}}{\delta s} \right){\rm d}x+\int_\Omega T\nabla \frac{\delta \mathsf{A}}{\delta\mathbf{m}}\cdot Q \cdot \nabla \frac{\delta \mathsf{B}}{\delta\mathbf{m}}{\rm d}x,
\end{align*}
which recovers the structure given in \cite{Mo1984b}.

\section{Parameterizing Thermodynamic Fluxes in terms of Thermodynamic Forces}\label{parameterizing-fluxes}

For the single and double generator bracket formulations, it remains to parameterize the thermodynamic fluxes ($\boldsymbol{\sigma}^{\rm fr}$, $\mathbf{j}_s$, $\mathbf{j}_i$, $j_i$) in terms of the thermodynamic forces ($\operatorname{Def}\mathbf{u}= \frac{1}{2}(\nabla\mathbf{u}+\nabla\mathbf{u}^\mathsf{T})$, $\nabla T$, $\nabla \mu_i$, $\mu_i$), such that $[\Smh,\Smh] = (\Smh,\Smh) \geq 0$ and therefore the second law of thermodynamics is satisfied. We start by defining the entropy generation rate $I$ by
\[
(\Smh,\Smh) = \int_\Omega I{\rm d}x
\]
and therefore, since $\sum_i\mathbf{j}_i=0$, $\sum_i j_i=0$, note that
\begin{equation}
\label{TI}
TI = J_\alpha X^\alpha =   \sigmafr : \nabla \mathbf{u}  -\mathbf{j}_s \cdot \nabla T -\sum_i  \mathbf{j}_i \cdot \nabla \mu_i- \sum_i j_i \mu_i,
\end{equation}
where $J_\alpha$ denotes the thermodynamic fluxes, $X^\alpha$ the thermodynamic forces. We then assume that thermodynamic fluxes are proportional to thermodynamic forces in the form
\[
J_\alpha = \sum_\beta L_{\alpha\beta} X^\beta
\]
where $L_{\alpha\beta}$ is a matrix of transport coefficients, that typically depends in a complicated fashion on the state variables. This has been found to be true for a wide range of irreversible processes \cite{dGMa1969}. If $L_{\alpha\beta}$ is a symmetric positive semi-definite matrix, then $J_\alpha X^\alpha= X^\beta L_{\alpha\beta}X^\alpha$ is a positive semi-definite quadratic form and therefore $TI \geq 0$. In fact, using Curie's principle \cite{dGMa1969}, there will be one matrix $L_{\alpha\beta}$ for each type of process: scalar, vector and tensor.

In determining the $L_{\alpha\beta}$ matrices, we will consider two distinct approaches. The first approach assumes that the thermodynamic fluxes represent the (molecular-scale) physical irreversible processes of viscous dissipation, heat conduction, diffusion and phase changes. This will lead to a set of equations suitable for the direct numerical simulation of geophysical fluid flows, and reduce to the well-known Navier-Stokes-Fourier equations in the case of a single component fluid. The second approach treats the thermodynamic fluxes as representing subgrid-scale turbulent fluxes, inspired by the approach in \cite{Gassmann2018,Gassmann2015} for the development of thermodynamically-consistent turbulent flux parameterizations. A powerful feature of the variational approach is the ability to treat both of these cases in a unified way.

\subsection{Parameterization of Physical Irreversible Processes}
We start by splitting $\sigmafr$ and $\text{Def } \uv$ into trace-free and scalar components as 
\begin{align*}
\sigmafr &= \sigmafrzero + \frac{1}{3} (\operatorname{Tr} \sigmafr) \delta \\
\text{Def } \uv &= (\text{Def } \uv)^{(0)} + \frac{1}{3} (\nabla \cdot \uv) \delta
\end{align*}
with unit diagonal tensor $\delta$ and where $\sigmafrzero$ and $(\text{Def } \uv)^{(0)}$ are trace-free. We will denote with $L_{ij}$ the matrix associated with vector processes, and with $\Ll_{ij}$ the matrix associated with scalar processes. Therefore we can write for the scalar processes (bulk viscosity, phase changes):
\[
\left[ \begin{matrix}
\operatorname{Tr} \sigmafr \\
-j_i \\
\dots \\
\end{matrix}
\right] =
\left[ \begin{matrix}
\Ll_{00} & \Ll_{0j} & \dots \\
\Ll_{i0} & \Ll_{ij} & \dots \\
\dots & \dots & \dots \\
\end{matrix}
\right]
\left[ \begin{matrix}
\frac{1}{3} \nabla \cdot \uv \\
\mu_j \\
\dots \\
\end{matrix}
\right],
\]
for the vector processes (heat conduction, diffusion):
\[
-\left[ \begin{matrix}
\mathbf{j}_s\\
\mathbf{j}_i \\
\dots \\
\end{matrix}
\right] =
\left[ \begin{matrix}
L_{ss} & L_{sj} & \dots \\
L_{is} & L_{ij} & \dots \\
\dots & \dots & \dots \\
\end{matrix}
\right]
\left[ \begin{matrix}
\nabla T \\
\nabla \mu_j \\
\dots \\
\end{matrix}
\right] ,
\]
and for the tensor process (shear viscosity):
\[
\sigmafrzero = 2 \mu (\text{Def }\uv)^{(0)},
\]
with $\mu \geq 0$ the shear viscosity coefficient. The off-diagonal elements represent cross effects, such as the Soret and Dufour effects in the vector case. Using this, we can write $\sigmafr$ as
\[
\sigmafr = 2 \mu \text{Def } \uv + \left(\frac{1}{9}\Ll_{00} - \frac{2}{3} \mu\right) \nabla \cdot \uv \delta + \frac{1}{3} \sum_i \Ll_{0i} \mu_i \delta.
\]
The transport coefficients $L_{\alpha\beta}$ and $\Ll_{\alpha\beta}$ must satisfy the Onsager-Casimir relations
\[
L_{si} = L_{is} \quad\quad L_{ij} = L_{ji} \quad\quad\text{and}\quad\quad \Ll_{0i} = -\Ll_{i0} \quad\quad \Ll_{ij} = \Ll_{ji}.
\]
Additionally, the mass control condition $\sum_i \ji = 0$ requires that
\[
\sum_i \Ll_{is} = \sum_i \Ll_{ij} = 0, \;\; \forall\;j
\]
and the mass control condition $\sum_i j_i = 0$ requires that
\[
\sum_i L_{i0} = \sum_i L_{ij} = 0,\;\; \forall \;j.
\]
The second law of thermodynamics and the Onsager-Casimir relationships ensure that $L_{\alpha\beta}$ and $\Ll_{\alpha\beta}$ are symmetric positive semi-definite and therefore $I \geq 0$. More details on this can be found in \cite{GayBalmaz2017}.

%Taken together, the mass control conditions and Onsager-Casimir relationships ensure that $L_{\alpha\beta}$ and $\Ll_{\alpha\beta}$ are symmetric positive semi-definite and therefore $I \geq 0$. More details on this can be found in \cite{GayBalmaz2017}. 

%A discussion of relationship of $I$, $L_{\alpha,\beta}$, $J_\alpha$ and $X_\alpha$ to dissipation functions and the GENERIC framework is found in Section \ref{generic-correspondence}.

\paragraph{Single component.} If there is a single component, the only irreversible processes are heat conduction and viscous dissipation, and $\mathbf{j}_i = j_i = 0$. Therefore, the thermodynamic fluxes are $\sigmafr$ and $\mathbf{j}_s$ and the thermodynamic forces are $\operatorname{Def}\mathbf{u}$ and $\nabla T$. There are no cross effects since there is only one of each type of process (again splitting viscous dissipation into shear viscosity and bulk viscosity). Therefore, we can write
\[
\sigmafr = 2 \mu \text{Def } \uv + (\zeta - \frac{2}{3} \mu) \nabla \cdot \uv \delta
\]
and
\[
\mathbf{j}_s = -\frac{\kappa }{T}\nabla T,
\]
where $\zeta = \frac{1}{9}\Ll_{00} \geq 0$ is the bulk viscosity coefficient and $\kappa = T L_{ss} \geq 0$ is the thermal conductivity. Stokes hypothesis $\zeta = 0$ is often employed, although the validity of this is for compressible flows is somewhat questionable.

\subsection{Parameterization of Turbulent Processes}
In the section, we treat the thermodynamic fluxes as representing turbulent subgrid fluxes arising from the closure of covariance terms in a mass-weighted Reynolds averaging, not true molecular scale, physical irreversible processes. This also implies an interpretation of predicted quantities as turbulence averaged. In doing so, we will follow the approach of \cite{Gassmann2018,Gassmann2015}. 

We start by rewriting $\nabla \mu_i$ and $\mathbf{j}_s$  as
\begin{align}
\label{mui-reformat}
\nabla \mu_i &= \nabla \mu_i|_T - \eta_i \nabla T \\
\label{js-reformat}
\js &= \frac{\mathbf{j}_s^h}{T} + \sum_i \eta_i \mathbf{j}_i,
\end{align}
where $\nabla \mu_i|_T$ is the gradient of $\mu_i(p,T,q_i)$ with $T$ held constant, $\eta_i = \pp{\eta}{q_i}(p,T,q_i)$ is the partial specific entropy and $\mathbf{j}_s^h = T(\mathbf{j}_s - \sum_i \eta_i \mathbf{j}_i)$ is the sensible heat flux. Using \eqref{mui-reformat} and \eqref{js-reformat}, we can rewrite the $TI$ equation \eqref{TI} as
\begin{equation}
\label{TI-reformat}
TI = \sigmafr : \nabla \mathbf{u} - \frac{\mathbf{j}_s^h}{T} \cdot \nabla T - \sum_i \mathbf{j}_i \cdot \nabla \mu_i|_T -\sum_i j_i \mu_i.
\end{equation}
Similarly, we can rewrite the entropy equation for the multicomponent compressible fluid (see \eqref{system_Eulerian_atmosphere_moist_L} or \eqref{system_Eulerian_atmosphere_moist_H}) as
\begin{equation}
\label{dSdt-reformat}
\bar D_t s+ \nabla \cdot \Big(\frac{\mathbf{j}_s^h}{T} + \sum_i \eta_i \mathbf{j}_i\Big) - \frac{1}{T} \sigmafr : \nabla \mathbf{u} + \frac{\mathbf{j}_s^h}{T^2} \cdot \nabla T + \frac{1}{T} \sum_i \mathbf{j}_i \cdot \nabla \mu_i|_T + \frac{1}{T} \sum_i  j_i \mu_i = 0.
\end{equation}
The thermodynamic fluxes are now ($\sigmafr$, $\mathbf{j}_s^h$, $\mathbf{j}_i$, $j_i$), and the thermodynamic forces are ($\operatorname{Def}\mathbf{u} $, $\frac{1}{T}\nabla T$, $\nabla \mu_i|_T$, $\mu_i$). 

\paragraph{Parameterization.} From the outset we will assume that $\sigmafr$ is trace-free, and therefore there are no cross-effects between viscous dissipation and phase changes. Therefore we can write for the scalar processes (phase changes):
\[
\left[ \begin{matrix}
-j_i \\
\dots \\
\end{matrix}
\right] =
\left[ \begin{matrix}
\Al_{i0} & \Al_{ij} & \dots \\
\dots & \dots & \dots \\
\end{matrix}
\right]
\left[ \begin{matrix}
\mu_j \\
\dots \\
\end{matrix}
\right] ,
\]
for the vector processes (heat conduction, diffusion):
\[
-\left[ \begin{matrix}
\mathbf{j}_s^h\\
\mathbf{j}_i \\
\dots \\
\end{matrix}
\right] =
\left[ \begin{matrix}
A_{ss} & A_{sj} & \dots \\
A_{is} & A_{ij} & \dots \\
\dots & \dots & \dots \\
\end{matrix}
\right]
\left[ \begin{matrix}
\frac{\nabla T}{T} \\
\nabla \mu_j|_T \\
\dots \\
\end{matrix}
\right] ,
\]
and for the tensor process (shear viscosity):
\[
\sigmafrzero =\rho \left[ \begin{matrix}
K^m_h E & K^m_h F & K^m_v G \\
K^m_h F & -K^m_h E & K^m_v H \\
K^m_v G & K^m_v H & 0 \\
\end{matrix} \right]
\]
with horizontal strain deformation $E = \partial_x u - \partial_y v$, horizontal shear deformation $F = \partial_x v + \partial_y u$, vertical strain deformation $G = \partial_z u + \partial_x w$ and vertical shear deformation $H=\partial_z v + \partial_y w$; where $K^m_h$ and $K^m_v$ are (positive) horizontal and vertical momentum diffusion coefficients. This is in fact the parameterization of \cite{Sm1993} adapted to the case of fully compressible flows. See also \cite{Becker2007,Gassmann2018a,Rolffs2013,Rolffs2018}.

Note that $\Al_{ij}$ is simply $\Ll_{ij}$ with the row and column corresponding to the trace of $\operatorname{Def}\mathbf{u}$ removed. As shown in \cite{GayBalmaz2017}, the matrices $A_{ij}$ and $L_{ij}$ are related by
\begin{equation}
A = M L M^T
\end{equation}
with
\begin{equation}
M = \left[ \begin{matrix}
T & -T \eta_1 & -T \eta_2 & -T \eta_3 & \dots \\
0 & 1 & 0 & 0 & \dots\\
0 & 0 & 1 & 0 & \dots\\
0 & 0 & 0 & 1 & \dots\\
\dots & \dots & \dots & \dots
\end{matrix}
\right].
\end{equation}
Since $M$ is invertible, $L_{ij}$ is symmetric positive semi-definitive if and only if $A_{ij}$ is symmetric positive semi-definite. The Onsager-Casimir relationships are then
\begin{equation}
A_{ij} = A_{ji} \quad\quad\text{and}\quad\quad \Al_{ij} = \Al_{ji}.
\end{equation}
Additionally, the mass control condition $\sum_i \ji = 0$ requires that
\begin{equation}
\sum_i \Al_{ij} = 0, \;\; \forall\;j
\end{equation}
and the mass control condition $\sum_i j_i = 0$ requires that
\begin{equation}
\sum_i A_{i0} = \sum_i A_{ij} = 0, \;\; \forall\;j.
\end{equation}

\paragraph{Choices made in \cite{Gassmann2015}.} Now we consider a fluid composed of four components: dry air ($d$), water vapor ($v$), liquid water condensate ($l$) and ice particles ($f$); and assume that $j_d = 0$ and $\mathbf{j}_l = \mathbf{j}_f = 0$. Therefore by the mass control conditions we have $j_v + j_l + j_f = 0$ and $\mathbf{j}_d + \mathbf{j}_v = 0$. We also make the further assumption that the fluxes $\mathbf{j}_s^h$, $\mathbf{j}_d$, $\mathbf{j}_v$ do not depend on the forces $\nabla \mu_l|_T$ and $\nabla \mu_f|_T$, i.e., we have $A_{sl}=A_{sf}=A_{il}=A_{if}=0$. This is slightly different to the assumption made in \cite{Gassmann2015}, where it was assumed that condensate and ice do not contribute to the pressure gradient i.e. $\nabla \mu_l|_T = \nabla \mu_f|_T = 0$. Our approach will give the same parameterization for the thermodynamic fluxes, but a slightly different form of the pressure gradient, although the difference in a numerical sense will be very small since condensate and ice are trace species. Additionally, if we assumed $\nabla \mu_l|_T = \nabla \mu_f|_T = 0$ then this would imply changes to the moist state equation, which would have implications for the treatment of phase change terms.

Taken together, these assumptions allow a further reduced form for the coefficient matrices obtained by eliminating some of the terms. We will eliminate $\mathbf{j}_d$ and $j_v$ in what follows. This gives finally for scalar processes (phase change):
\[
\left[ \begin{matrix}
-j_l \\
-j_f \\
\end{matrix}
\right] =
\left[ \begin{matrix}
\Al_{ll} & \Al_{lf} \\
\Al_{fl} & \Al_{ff} \\
\end{matrix}
\right]
\left[ \begin{matrix}
\mu_l - \mu_v \\
\mu_f - \mu_v \\
\end{matrix}
\right] 
\]
and for the vector processes (heat conduction, diffusion):
\[
\-\left[ \begin{matrix}
\mathbf{j}_s^h\\
\mathbf{j}_v \\
\end{matrix}
\right] =
\left[ \begin{matrix}
A_{ss} & A_{sv} \\
A_{vs} & A_{vv} \\
\end{matrix}
\right]
\left[ \begin{matrix}
\frac{\nabla T}{T} \\
\nabla(\mu_v - \mu_d)|_T \\
\end{matrix}
\right].
\]
Now we assume parameterizations for the vector processes of the form
\[
-\mathbf{j}_s^h = A_{ss}\cdot \frac{1}{T} \nabla T= c_p \rho K^h \cdot \nabla T
\]
and
\[
- \mathbf{j}_v = A_{vv}\cdot \nabla (\mu_v - \mu_d)|_T = \frac{\rho^{2} q_v q_d}{p} K^v \cdot \nabla (\mu_v - \mu_d)|_T,
\]
where $K^h$ and $K^v$ are tensors with only diagonal components, to allow a distinction between horizontal and vertical mixing. This makes physical sense, since the fluid is stratified and the grid resolution is well above the Ozmidov scale of isotropic turbulence. These parameterizations are exactly what is done in equations (32) and (46) in \cite{Gassmann2015}, and lead to a diagonal positive semi-definite $A_{\alpha\beta}$ matrix (which will give a non-diagonal but still positive semi-definite $L_{\alpha\beta}$ matrix, which shows the somewhat hidden cross-effects). For the scalar processes, we simply follow \cite{Gassmann2015} and note that the coefficients $\Al_{\alpha\beta}$ are determined by the microphysics scheme in such a way that $j_l$ ($j_f$) has an opposite sign to $\mu_l - \mu_v$ ($\mu_f - \mu_v$), which ensures that $TI$ is positive.

\paragraph{Identification of variables.} Now making the formal identifications
\[
\begin{array}{ccccccccc}
T &\rightarrow& \hat{T} &\qquad I &\rightarrow& \sigma& \qquad \sigmafr : \nabla \uv &\rightarrow& \varepsilon_{sh} \\
\mathbf{j}_s^h &\rightarrow& \mathbf{J}_s& \qquad \mathbf{j}_i &\rightarrow& \mathbf{J}_i^*& \qquad \mu_i &\rightarrow& \hat{\mu}_i \\
j_i &\rightarrow& I_i& \qquad \eta_i &\rightarrow& \hat{s}_i &\qquad q_i &\rightarrow& \hat{q}_i \\
\rho &\rightarrow& \bar{\rho}& \qquad \mathbf{u} &\rightarrow& \hat{\mathbf{v}}& \qquad \sigmafr &\rightarrow& - \overline{\rho v^{\prime\prime} v^{\prime\prime}} \\
\eta &\rightarrow& \hat{s}
\end{array}
\]
it is easy to see that \eqref{TI-reformat} is equivalent to equation (28) from \cite{Gassmann2015} and \eqref{dSdt-reformat} is equivalent to equation (20), when we assume $\mathbf{J}_i^d = 0$ (no precipitation, which is what is done in this paper). These are the two fundamental equations of \cite{Gassmann2015}. Finally we obtain that the system of equations (1)--(4) from \cite{Gassmann2015} is equivalent to the system \eqref{system_Eulerian_atmosphere_moist_L_final} (see also the equivalent systems written in Appendix \ref{alternative-prognostic}). This is a demonstration that the variational and bracket formulations reproduce the equations in \cite{Gassmann2015}.  It is believed that the revised formulation in \cite{Gassmann2018} can also be written as a bracket system, but the demonstration of this is left to future work. 

\section{Conclusions}
\label{conclusions}
This paper has presented bracket formulations for multicomponent, multiphase fully compressible geophysical fluids undergoing reversible and irreversible processes (viscous dissipation, heat conduction, diffusion, phase changes), based on a variational principle that incorporates irreversible processes \cite{GayBalmaz2017,GBYo2017b}. It was shown that many different prior approaches in the literature fit into this framework, including the single generator bracket \cite{BeEd1994,EdBe1991b,EdBe1991a} and the double generator bracket \cite{Ka1984}, as well as its metriplectic (or GENERIC) specific cases \cite{GrOtt1997,Mo1986,OtGr1997}. All these bracket formulations can be systematically derived in a constructive way from the general set of equations obtained via the variational principle, whereas so far the derivation of bracket formulations was mainly empirical.

A key element is the parameterization of the thermodynamic fluxes ($\boldsymbol{\sigma}^{\rm fr}$, $\mathbf{j}_s$, $\mathbf{j}_i$, $j_i$) in terms of the thermodynamic forces ($\operatorname{Def}\mathbf{u}= \frac{1}{2}(\nabla\mathbf{u}+\nabla\mathbf{u}^\mathsf{T})$, $\nabla T$, $\nabla \mu_i$, $\mu_i$). The first law of thermodynamics (conservation of energy) is satisfied independent of the choice of parameterization, while the second law requires that the parameterizations obey certain rules. In particular, ensuring that the Onsager-Casimirs relationships and mass control conditions are satisfied gives positive-definite entropy production. Two different parameterizations were presented: an approach treating thermodynamic fluxes as physical irreversible processes, that reduces to the Navier-Stokes-Fourier equations for a single component; and an approach treating them as subgrid turbulence parameterizations that yields the equations of \cite{Gassmann2015}.

The main limitations of this work are the assumptions of a single temperature and velocity for all components; and of a materially and adiabatically closed domain. These limitations will be removed in future work through the extension of the formulation to fluids with open boundaries and multiple temperatures and velocities. This (along with the incorporation of chemistry and radiation) is necessary to treat the irreversible processes of  precipitation, chemical reactions and radiation; and to handle boundary fluxes of mass, entropy and energy. It will also enable a rigorous examination of the conditions under which the simplifying assumptions of a single temperature and velocity (perhaps with a separate, constant fall velocity for hydrometeors) are valid.

The treatment of thermodynamic fluxes as subgrid turbulence parameterizations is also an area ripe for improvement. In particular, the approach outlined here does not incorporate any memory into the subgrid processes: for example resolved kinetic energy dissipated by the viscous term in the velocity equation is instantaneously transformed into heat. It also cannot treat non-local subgrid processes such as convection, gravity wave drag and boundary layer processes. We aim to extend this formulation to overcome these limitations, through the incorporation of more sophisticated treatments of the subgrid processes such as  conditional filtering \cite{Thuburn2018b,Thuburn2018a}, Lagrangian averaging \cite{Gilbert2018,Holm2002a}, eddy diffusivity mass flux \cite{Tan2018,Yano2012}, convected fluid microstructure \cite{Holm2018,Holm2012} or stochastic Lagrangian averaged transport (SALT) \cite{Cotter2017,Drivas2018,GBHo2018b,GBHo2018a,Holm2015}.

It is also planned to study the numerical implementation of these equations, in particular in a way that preserves the key elements of the dissipation bracket structure in both space and time, following existing work \cite{EldredDubos2018} done for the reversible dynamics using the Hamiltonian formulation, compatible Galerkin methods and Poisson time integrators. In fact, this has already been done using finite-differences for the spatial discretization in \cite{Gassmann2013,Gassmann2018,Gassmann2008,Gassmann2015}, and we aim to extend this work to compatible Galerkin methods and metriplectic time integrators.

Other possible future work could include: variational and bracket formulations of semi-compressible fluids (Boussinesq, anelastic, pseudo-incompressible, semi-hydrostatic), non-Eulerian vertical coordinates, and the study of energy-Casimir theory for the metriplectic system.

\section{Acknowledgements}
Christopher Eldred was supported by the French National Research Agency through contract ANR-14-CE23-0010 (HEAT).

\bibliographystyle{abbrv}
\bibliography{main}

\begin{thebibliography}{10}

\bibitem{Bannon2002}
P.~R. Bannon.
\newblock Theoretical foundations for models of moist convection.
\newblock {\em Journal of the Atmospheric Sciences}, 52:1967--1982, 2002.

\bibitem{Bannon2003}
P.~R. Bannon.
\newblock Hamiltonian description of idealized binary geophysical fluids.
\newblock {\em Journal of the Atmospheric Sciences}, 60(22):2809--2819, 2003.

\bibitem{BaGB2018}
W.~Bauer and F.~{Gay-Balmaz}.
\newblock Towards a variational discretization of compressible fluidsfluids:
  the rotating shallow water equations.
\newblock {\em J. Comp. Dyn., accepted}, https://arxiv.org/pdf/ 1711.10617.pdf,
  2018.

\bibitem{Becker2007}
E.~Becker and U.~Burkhardt.
\newblock Nonlinear horizontal diffusion for {GCM}s.
\newblock {\em Monthly Weather Review}, 135(4):1439--1454, 2007.

\bibitem{BeEd1994}
A.~N. Beris and B.~J. Edwards.
\newblock {\em Thermodynamics of Flowing Systems with Internal Microstructure}.
\newblock Oxford University Press, 1994.

\bibitem{BrBaBiGBML2018}
R.~Brecht, W.~Bauer, A.~Bihlo, F.~{Gay-Balmaz}, and S.~MacLachlan.
\newblock Variational integrator for the rotating shallow-water equations on
  the sphere.
\newblock {\em ArXiv e-prints}, https://arxiv.org/pdf/1808.10507.pdf, 2018.

\bibitem{Br1970}
F.~P. Bretherton.
\newblock A note on {H}amilton's principle for perfect fluids.
\newblock {\em J. Fluid Mech.}, 44:19--31, 1970.

\bibitem{Charron2018a}
M.~Charron and A.~Zadra.
\newblock Hidden symmetries, trivial conservation laws and casimir invariants
  in geophysical fluid dynamics.
\newblock {\em Journal of Physics Communications}, 2(11):115018, 2018.

\bibitem{Charron2018}
M.~Charron and A.~Zadra.
\newblock On the triviality of potential vorticity conservation in geophysical
  fluid dynamics.
\newblock {\em Journal of Physics Communications}, 2(7):075003, 2018.

\bibitem{Cotter2017}
C.~J. Cotter, G.~A. Gottwald, and D.~D. Holm.
\newblock Stochastic partial differential fluid equations as a diffusive limit
  of deterministic lagrangian multi-time dynamics.
\newblock {\em Proceedings of the Royal Society of London A: Mathematical,
  Physical and Engineering Sciences}, 473(2205), 2017.

\bibitem{Cr1991}
G.~Craig.
\newblock A three-dimensional generalization of {E}liassen's balanced vortex
  equations derived from hamilton's principle.
\newblock {\em Quarterly Journal of the Royal Meteorological Society},
  117:435--448, 1991.

\bibitem{dGMa1969}
S.~de~Groot and P.~Mazur.
\newblock {\em Nonequilibrium Thermodynamics}.
\newblock North-Holland, 1969.

\bibitem{DeSa2005}
P.~J. Dellar and R.~Salmon.
\newblock Shallow water equations with a complete {C}oriolis force and
  topography.
\newblock {\em Phys. Fluids}, 17(106), 2005.

\bibitem{DeGaGBZe2014}
M.~Desbrun, E.~Gawlik, F.~Gay-Balmaz, and V.~Z. Zeitlin.
\newblock Variational discretization for rotating stratified fluids.
\newblock {\em Disc. Cont. Dyn. Syst. Series A}, 32(2):479--511, 2014.

\bibitem{Drivas2018}
T.~D. Drivas and D.~D. Holm.
\newblock Circulation and energy theorem preserving stochastic fluids.
\newblock {\em arXiv preprint arXiv:1808.05308}, 2018.

\bibitem{Dubos2015}
T.~Dubos, S.~Dubey, M.~Tort, R.~Mittal, Y.~Meurdesoif, and F.~Hourdin.
\newblock Dynamico-1.0, an icosahedral hydrostatic dynamical core designed for
  consistency and versatility.
\newblock {\em Geoscientific Model Development}, 8(10):3131--3150, 2015.

\bibitem{Dubos2014}
T.~Dubos and M.~Tort.
\newblock Equations of atmospheric motion in non-{E}ulerian vertical
  coordinates: Vector-invariant form and quasi-{H}amiltonian formulation.
\newblock {\em Monthly Weather Review}, 142(10):3860--3880, 2014.

\bibitem{Ec1960}
C.~Eckart.
\newblock Variation principles of hydrodynamics.
\newblock {\em Phys. Fluids}, 3:421--427, 1960.

\bibitem{Ed1998}
B.~J. Edwards.
\newblock An analysis of single and double generator thermodynamic formalisms
  for complex fluids.
\newblock {\em J. Non-Equilib. Thermodyn.}, 23:301--333, 1998.

\bibitem{EdBe1991b}
B.~J. Edwards and A.~N. Beris.
\newblock Noncanonical poisson bracket for nonlinear elasticity with extensions
  to viscoelasticity.
\newblock {\em Phys. A: Math. Gen.}, 24:2461--2480, 1991.

\bibitem{EdBe1991a}
B.~J. Edwards and A.~N. Beris.
\newblock Unified view of transport phenomena based on the generalized bracket
  formulation.
\newblock {\em Ind. Eng. Chem. Res.}, 30:873--881, 1991.

\bibitem{EdBeOt1998}
B.~J. Edwards, A.~N. Beris, and H.-C. \"Ottinger.
\newblock An analysis of single and double generator thermodynamic formalisms
  for complex fluids. {II}. {T}he microscopic description.
\newblock {\em J. Non-Equilib. Thermodyn.}, 23:334--350, 1998.

\bibitem{EldredDubos2018}
C.~Eldred, T.~Dubos, and E.~Kritsikis.
\newblock A quasi-{H}amiltonian discretization of the thermal shallow water
  equations.
\newblock {\em Journal of Computational Physics}, 2018.

\bibitem{Eldred2017}
C.~Eldred and D.~Randall.
\newblock Total energy and potential enstrophy conserving schemes for the
  shallow water equations using hamiltonian methods -- part 1: Derivation and
  properties.
\newblock {\em Geoscientific Model Development}, 10(2):791--810, 2017.

\bibitem{Gassmann2013}
A.~Gassmann.
\newblock A global hexagonal {C}-grid non-hydrostatic dynamical core
  ({ICON-IAP}) designed for energetic consistency.
\newblock {\em Quarterly Journal of the Royal Meteorological Society},
  139(670):152--175, 2013.

\bibitem{Gassmann2018a}
A.~Gassmann.
\newblock Discretization of generalized {C}oriolis and friction terms on the
  deformed hexagonal {C}-grid.
\newblock {\em Quarterly Journal of the Royal Meteorological Society},
  144(716):2038--2053, 2018.

\bibitem{Gassmann2018}
A.~Gassmann.
\newblock Entropy production due to subgrid-scale thermal fluxes with
  application to breaking gravity waves.
\newblock {\em Quarterly Journal of the Royal Meteorological Society},
  144(711):499--510, 2018.

\bibitem{Gassmann2008}
A.~Gassmann and H.-J. Herzog.
\newblock Towards a consistent numerical compressible non-hydrostatic model
  using generalized {H}amiltonian tools.
\newblock {\em Quarterly Journal of the Royal Meteorological Society},
  134(635):1597--1613, 2008.

\bibitem{Gassmann2015}
A.~Gassmann and H.-J. Herzog.
\newblock How is local material entropy production represented in a numerical
  model?
\newblock {\em Quarterly Journal of the Royal Meteorological Society},
  141(688):854--869, 2015.

\bibitem{GayBalmaz2017}
F.~{Gay-Balmaz}.
\newblock A variational derivation of the thermodynamics of a moist atmosphere
  with rain process and its pseudoincompressible approximation.
\newblock {\em ArXiv e-prints}, https://arxiv.org/pdf/1701.03921.pdf, 2018.

\bibitem{GBHo2013}
F.~{Gay-Balmaz} and D.~D. Holm.
\newblock Selective decay by {C}asimir dissipation in inviscid fluids.
\newblock {\em Nonlinearity}, 26:495--524, 2013.

\bibitem{GBHo2014}
F.~{Gay-Balmaz} and D.~D. Holm.
\newblock A geometric theory of selective decay with applications in {MHD}.
\newblock {\em Nonlinearity}, 27:1747--1777, 2014.

\bibitem{GBHo2018b}
F.~{Gay-Balmaz} and D.~D. Holm.
\newblock Predicting uncertainty in geometric fluid mechanics.
\newblock {\em Disc. Cont. Dyn. Syst. Series S.}, to appear, 2018.

\bibitem{GBHo2018a}
F.~{Gay-Balmaz} and D.~D. Holm.
\newblock Stochastic geometric models with non-stationary spatial correlations
  in {L}agrangian fluid flows.
\newblock {\em J. Nonlin. Sci.}, 28(3):873--904, 2018.

\bibitem{GBYo2017a}
F.~{Gay-Balmaz} and H.~Yoshimura.
\newblock A {L}agrangian variational formalism for nonequilibrium
  thermodynamics. {P}art {I}: discrete systems.
\newblock {\em J. Geom. Phys.}, 111:169--193, 2017.

\bibitem{GBYo2017b}
F.~{Gay-Balmaz} and H.~Yoshimura.
\newblock A {L}agrangian variational formalism for nonequilibrium
  thermodynamics. {P}art {II}: continuum systems.
\newblock {\em J. Geom. Phys.}, 111:194--212, 2017.

\bibitem{GBYo2018}
F.~{Gay-Balmaz} and H.~Yoshimura.
\newblock A variational formulation of nonequilibrium thermodynamics for
  discrete open systems with mass and heat transfer.
\newblock {\em Entropy}, 3:163, 2018.

\bibitem{Gilbert2018}
A.~D. Gilbert and J.~Vanneste.
\newblock Geometric generalised {L}agrangian-mean theories.
\newblock {\em Journal of Fluid Mechanics}, 839:95--134, 2018.

\bibitem{Gr1984}
M.~Grmela.
\newblock Bracket formulation of dissipative fluid mechanics equations.
\newblock {\em Phys. Lett. A}, 102:355--358, 1984.

\bibitem{GrOtt1997}
M.~Grmela and H.-C. \"Ottinger.
\newblock Dynamics and thermodynamics of complex fluids. {I}. {D}evelopment of
  a general formalism.
\newblock {\em Phys. Rev. E}, 56:6620--6632, 1997.

\bibitem{He1955}
J.~W. Herivel.
\newblock The derivation of the equations of motion of an ideal fluid by
  {H}amilton's principle.
\newblock {\em Proc. Cambridge Philos. Soc.}, 51:344--349, 1955.

\bibitem{Ho1996}
D.~D. Holm.
\newblock {H}amiltonian balance equations.
\newblock {\em Physica D}, 98(2):379--414, 1996.

\bibitem{Holm2002a}
D.~D. Holm.
\newblock {L}agrangian averages, averaged {L}agrangians, and the mean effects
  of fluctuations in fluid dynamics.
\newblock {\em Chaos: An Interdisciplinary Journal of Nonlinear Science},
  12(2):518--530, 2002.

\bibitem{Holm2015}
D.~D. Holm.
\newblock Variational principles for stochastic fluid dynamics.
\newblock {\em Proceedings of the Royal Society of London A: Mathematical,
  Physical and Engineering Sciences}, 471(2176), 2015.

\bibitem{Holm2018}
D.~D. Holm.
\newblock Stochastic parametrization of the richardson triple.
\newblock {\em Journal of Nonlinear Science}, Jun 2018.

\bibitem{HoMaRa2002}
D.~D. Holm, J.~E. Marsden, and T.~S. Ratiu.
\newblock The {E}uler-{P}oincar\'e equations in geophysical fluid dynamics.
\newblock {\em Large-scale atmosphere-ocean dynamics}, Vol. II:251--300, 2002.

\bibitem{HoMaRaWe1985}
D.~D. Holm, J.~E. Marsden, T.~S. Ratiu, and A.~Weinstein.
\newblock Nonlinear stability of fluid and plasma equilibria.
\newblock {\em Phys. Rep.}, 123:1--116, 1985.

\bibitem{Holm2012}
D.~D. Holm and C.~Tronci.
\newblock Multiscale turbulence models based on convected fluid microstructure.
\newblock {\em Journal of Mathematical Physics}, 53(11):115614, 2012.

\bibitem{Ka1984}
A.~Kaufman.
\newblock Dissipative {H}amiltonian systems: A unifying principle.
\newblock {\em Phys. Lett. A}, 100:419--422, 1984.

\bibitem{Li1963}
C.~C. Lin.
\newblock Liquid helium.
\newblock {\em Proc. Int. School of Physics}, Course XXI:421--427, 1960.

\bibitem{MaRaWe1984}
J.~E. Marsden, T.~S. Ratiu, and A.~Weinstein.
\newblock Semidirect product and reduction in mechanics.
\newblock {\em Trans. Amer. Math. Soc.}, 281:147--177, 1984.

\bibitem{MaWe1983}
J.~E. Marsden and A.~Weinstein.
\newblock {C}oadjoint orbits, vortices, and {C}lebsch variables for
  incompressible fluids.
\newblock {\em Physica D: Nonlinear Phenomena}, 7(1):305--323, 1983.

\bibitem{MiSa1985}
J.~Miles and R.~Salmon.
\newblock Weakly dispersive nonlinear gravity waves.
\newblock {\em J. Fluid Mech.}, 157:519--531, 1985.

\bibitem{Mo1984a}
P.~Morrison.
\newblock Bracket formulation for irreversible classical elds.
\newblock {\em Phys. Lett. A}, 100:423--427, 1984.

\bibitem{Mo1984b}
P.~Morrison.
\newblock Some observations regarding brackets and dissipation.
\newblock Technical report, University of California, Berkeley, 1984.

\bibitem{Mo1986}
P.~Morrison.
\newblock A paradigm for joined hamiltonian and dissipative systems.
\newblock {\em Physica D}, 18:410--419, 1986.

\bibitem{MoGr1980}
P.~Morrison and J.~Greene.
\newblock Noncanonical {H}amiltonian density formulation of hydrodynamics and
  ideal magnetohydrodynamics.
\newblock {\em Phys. Rev. Letters}, 45:790--794, 1980.

\bibitem{Ol2006}
M.~Oliver.
\newblock Variational asymptotics for rotating shallow water near geostrophy: a
  transformational approach.
\newblock {\em J. Fluid Mech.}, 551:197--234, 2006.

\bibitem{OtGr1997}
H.-C. \"Ottinger and M.~Grmela.
\newblock Dynamics and thermodynamics of complex fluids. {II.} {I}llustrations
  of a general formalism.
\newblock {\em Phys. Rev. E}, 56:6633--6655, 1997.

\bibitem{PaMuToKaMaDe2010}
D.~Pavlov, P.~Mullen, Y.~Tong, E.~Kanso, and J.~E. Marsden.
\newblock Structure-preserving discretization of incompressible fluids.
\newblock {\em Physica D}, 240:443--458, 2010.

\bibitem{Ri1981}
P.~Ripa.
\newblock Symmetries and conservation laws for internal gravity waves.
\newblock In {\em Nonlinear Properties of Internal Waves}, volume~76, pages
  281--306. AIP Publishing, New York, 1981.

\bibitem{Sa1983}
R.~Salmon.
\newblock Practical use of {H}amilton's principle.
\newblock {\em J. Fluid Mech.}, 132:431--44, 1983.

\bibitem{Sa1985}
R.~Salmon.
\newblock New equations for nearly geostrophic flow.
\newblock {\em J. Fluid Mech.}, 153:461--477, 1985.

\bibitem{Sa1988}
R.~Salmon.
\newblock Hamilton fluid dynamics.
\newblock {\em Ann. Rev. Fluid Mech.}, 20:225--256, 1988.

\bibitem{Salmon2004}
R.~Salmon.
\newblock Poisson-bracket approach to the construction of energy- and
  potential-enstrophy-conserving algorithms for the shallow-water equations.
\newblock {\em Journal of the Atmospheric Sciences}, 61(16):2016--2036, 2004.

\bibitem{Rolffs2013}
U.~Schaefer-Rolffs and E.~Becker.
\newblock Horizontal momentum diffusion in {GCM}s using the dynamic
  {S}magorinsky model.
\newblock {\em Monthly Weather Review}, 141(3):887--899, 2013.

\bibitem{Rolffs2018}
U.~Schaefer-Rolffs and E.~Becker.
\newblock Scale-invariant formulation of momentum diffusion for high-resolution
  atmospheric circulation models.
\newblock {\em Monthly Weather Review}, 146(4):1045--1062, 2018.

\bibitem{ScHeGa2003}
T.~Schneider, I.~M. Held, and S.~T. Garner.
\newblock Boundary effects in potential vorticity dynamics.
\newblock {\em Journal of the Atmospheric Sciences}, 60:1024, 2003.

\bibitem{SeWh1968}
R.~L. Seliger and G.~B. Whitham.
\newblock Variational principles in continuum mechanics.
\newblock {\em Proc. Roy. Soc. A.}, 305:1--25, 1968.

\bibitem{Se1959}
J.~Serrin.
\newblock Mathematical principles of classical fluid mechanics.
\newblock {\em Handbuch der Physik VIII-I}, 51:125--263, 1959.

\bibitem{Shepherd1990}
T.~G. Shepherd.
\newblock Symmetries, conservation laws, and hamiltonian structure in
  geophysical fluid dynamics.
\newblock volume~32 of {\em Advances in Geophysics}, pages 287 -- 338.
  Elsevier, 1990.

\bibitem{Shepherd1993}
T.~G. Shepherd.
\newblock A unified theory of available potential energy.
\newblock {\em Atmosphere-Ocean}, 31(1):1--26, 1993.

\bibitem{Sm1993}
J.~Smagorinsky.
\newblock Some historical remarks on the use of nonlinear viscosities.
\newblock In B.~Galperin and S.~A. Orszag, editors, {\em Large Eddy Simulation
  of Complex Engineering and Geophysical Flows}, pages 3--36. Cambridge
  University Press: Cambridge, UK, 1993.

\bibitem{Tan2018}
Z.~Tan, C.~M. Kaul, K.~G. Pressel, Y.~Cohen, T.~Schneider, and J.~Teixeira.
\newblock An extended eddy-diffusivity mass-flux scheme for unified
  representation of subgrid-scale turbulence and convection.
\newblock {\em Journal of Advances in Modeling Earth Systems}, 10(3):770--800,
  2018.

\bibitem{Thuburn2018b}
J.~Thuburn and G.~K. Vallis.
\newblock Properties of conditionally filtered equations: Conservation, normal
  modes, and variational formulation.
\newblock {\em Quarterly Journal of the Royal Meteorological Society},
  144(714):1555--1571, 2018.

\bibitem{Thuburn2018a}
J.~Thuburn, H.~Weller, G.~K. Vallis, R.~J. Beare, and M.~Whitall.
\newblock A framework for convection and boundary layer parameterization
  derived from conditional filtering.
\newblock {\em Journal of the Atmospheric Sciences}, 75(3):965--981, 2018.

\bibitem{Tort2014a}
M.~Tort and T.~Dubos.
\newblock Usual approximations to the equations of atmospheric motion: A
  variational perspective.
\newblock {\em Journal of the Atmospheric Sciences}, 71(7):2452--2466, 2014.

\bibitem{Tort2015}
M.~Tort, T.~Dubos, and T.~Melvin.
\newblock Energy-conserving finite-difference schemes for quasi-hydrostatic
  equations.
\newblock {\em Quarterly Journal of the Royal Meteorological Society},
  141(693):3056--3075, 2015.

\bibitem{Yano2012}
J.-I. Yano.
\newblock Mass-flux subgrid-scale parameterization in analogy with
  multi-component flows: a formulation towards scale independence.
\newblock {\em Geoscientific Model Development}, 5(6):1425--1440, 2012.

\end{thebibliography}

\appendix

\section{Curl-Form Formulations}
\label{alternative-prognostic}
Here we present some alternative choices of prognostic variables: $(\vv,\rho_i,s)$, $(\vv,\rho_i,\eta)$, $(\vv,\rho,q_k,s)$ and $(\vv,\rho,q_k,\eta)$; recalling that $i = 1,\dots,n$ sums over all the components and $k=1,\dots,n-1$ sums over the sparse components. These alternatives are generally referred to as curl-form formulations, since the absolute velocity $\vv$ is predicted instead of absolute momentum $\mv$, and this leads to the appearance of a term involving the curl of $\vv$. It would also be possible to replace some of the $\rho_i$ with $q_i$ or some of the $q_k$ with $\rho_k$; or use an alternative Lie-Poisson formulation that uses $(\rho,\rho_k,\mv,s)$ (as done for a binary fluid in \cite{Bannon2003}), but these choices are not discussed further, as they can be easily obtained by a change of variables similar to those we describe below. Functionals using curl-form variables are denoted as $\Ah$ instead of $\Amh$. The $(\vv,\rho_i,s)$ variant is discussed in detail in the following section, while for the others only the main results (chain rule, Hamiltonian, Poisson Bracket, Dissipation brackets and equations of motion) are given.

\subsection{Variational Formulation for $(\vv,\rho_i,s)$}
We start with the variational formulation based on $\vv = \frac{1}{\rho} \frac{\delta \mathcal{L}}{\delta \mathbf{u}}$ instead of $\mathbf{m} = \frac{\delta \mathcal{L}}{\delta \mathbf{u}}$. Using the continuity equation $\partial_t\rho+\nabla\cdot (\rho\,\mathbf{u})=0$ gives the following form of equation \eqref{EP-m}
\begin{equation}
\label{EP-v}
\partial_t \left( \frac{1}{\rho} \frac{\delta\mathcal{L}}{\delta \mathbf{u}} \right) + \mathsf{L}_\mathbf{u}\left( \frac{1}{\rho}\frac{\delta\mathcal{L}}{\delta \mathbf{u}} \right) - \sum_i q _i \nabla \frac{\delta\mathcal{L}}{\delta\rho_i} - \eta \nabla \frac{\delta\mathcal{L}}{\delta s} = 0.
\end{equation}
The second term can be expanded using the Lie derivative expression $\mathsf{L}_\mathbf{u} \mathbf{v} = (\nabla \times \mathbf{v}) \times \uv + \nabla (\uv \cdot \mathbf{v})$ for one-forms, to get the equation in curl-form\footnote{These are termed curl-form due to the appearance of the $\nabla \times \left( \frac{1}{\rho}\frac{\delta\mathcal{L}}{\delta \mathbf{u}}\right)$ term} as
\begin{equation}
\label{EP-v-curl}
\partial_t \left( \frac{1}{\rho} \frac{\delta\mathcal{L}}{\delta \mathbf{u}} \right) + \nabla \times \left( \frac{1}{\rho}\frac{\delta\mathcal{L}}{\delta \mathbf{u}}\right) \times \uv + \nabla \left( \uv \cdot \frac{1}{\rho}\frac{\delta\mathcal{L}}{\delta \mathbf{u}} \right) -  \sum_i q _i \nabla \frac{\delta\mathcal{L}}{\delta\rho_i} - \eta \nabla \frac{\delta\mathcal{L}}{\delta s} = 0.
\end{equation}
Introducing now
\begin{align}
\label{bi-fv-defs}
\vv := \frac{1}{\rho} \frac{\delta \mathcal{L}}{\delta \mathbf{u}},\qquad B_i :=  \uv \cdot \frac{1}{\rho}\frac{\delta\mathcal{L}}{\delta \mathbf{u}} - \frac{\delta\mathcal{L}}{\delta \rho_i} = \uv \cdot \vv - \frac{\delta\mathcal{L}}{\delta \rho_i}, \qquad
\Fv := \rho \mathbf{u}, \qquad T := -\frac{\delta\mathcal{L}}{\delta s},
\end{align}
equations \eqref{EP-v-curl}, \eqref{rho-kinematic}, \eqref{S-kinematic},  can be rewritten as
\begin{align}
\label{v-eqn-EPv}
&\partial _t\vv  + \frac{\nabla \times \vv}{\rho} \times \Fv + \sum_i q _i \nabla B_i + \eta \nabla T = 0 \\
\label{rho-eqn-EPv}
&\partial _t\rho_i + \nabla \cdot (q _i \Fv)=0 \textcolor{white}{\sum_i }\\
\label{S-eqn-EPv}
&\partial _ts + \nabla \cdot (\eta \Fv) =0.
\end{align}
We have used the same symbol $B_i$ before. These equations are naturally connected, on the Hamiltonian side, to the curl-form Poisson formulation. The specific Lagrangian \eqref{specific-lagrangian} gives
\begin{align}
\label{specific-bi}
B_i &= K + \Phi + \mu_i, \qquad \vv = \rho \uv + \rho \Rv.
\end{align}

\subsection{Hamiltonian Formulation for $(\vv,\rho_i,s)$}

\paragraph{Chain rule.} Writing $\Ah[\vv,\rho_i,s] = \Amh[\mv,\rho_i,s]$ for an arbitrary functional, the chain rule for functional derivatives gives
\begin{align}
\label{chain-rule-vv-rhoi-s}
\frac{\delta \mathcal{A}}{\delta \rho_i} =\frac{\delta \mathsf{A}}{\delta \rho_i} + \mathbf{v} \cdot \frac{\delta \mathsf{A}}{\delta\mathbf{m}}, \qquad
\frac{\delta \mathcal{A}}{\delta \mathbf{v}} = \rho \dede{\Amh}{\mv},  \qquad
\frac{\delta \mathcal{A}}{\delta  s} =\frac{\delta \mathsf{A}}{\delta s}.
\end{align}
This can be used to transform the Lie-Poisson bracket \eqref{lie-poisson-bracket}.

\paragraph{Hamiltonian function.} We have $\Hh[\vv,\rho_i,s] = \Hmh[\mv,\rho_i,s]$ and therefore, using the chain rule \eqref{chain-rule-vv-rhoi-s} in \eqref{dHdx},
\begin{align}
\label{dHdx-vv}
\frac{\delta \mathcal{H}}{\delta \mathbf{v}} := \Fv = \rho \mathbf{u}, \qquad
\frac{\delta \mathcal{H}}{\delta \rho_i} := B_i = \uv \cdot \vv - \frac{\delta \mathcal{L}}{\delta \rho_i}, \qquad
\frac{\delta \mathcal{H}}{\delta s} := - \frac{\delta \mathcal{L}}{\delta s}.
\end{align}
These fit with \eqref{bi-fv-defs}. Using the specific Lagrangian (\ref{specific-lagrangian}) this gives
\begin{align}
\frac{\delta \mathcal{H}}{\delta \mathbf{v}} := \Fv =\rho \mathbf{u}, \qquad
\frac{\delta \mathcal{H}}{\delta \rho_i}:= B_i = K + \Phi + \mu_i, \qquad
\frac{\delta \mathcal{H}}{\delta s}:= T 
\end{align}
which fit with \eqref{specific-bi} and \eqref{bi-fv-defs}.

\paragraph{The Lie-Poisson bracket in curl-form.} By using the chain rule \eqref{chain-rule-vv-rhoi-s} in the Lie-Poisson bracket \eqref{M-m-bracket}--\eqref{S-m-bracket}, we obtain the curl-form Poisson bracket for functionals $\mathcal{A}[\vv,\rho_i,s]$, $\mathcal{B}[\vv,\rho_i,s]$, expressed as the sum of $2 + n$ brackets
\begin{equation}
\{\Ah,\Bh\} =  \{\Ah,\Bh\}_Q + \sum_i \{\Ah,\Bh\}_{R_i} +\{\Ah,\Bh\}_S,
\end{equation}
where the three terms are
\begin{align}
\label{Q-v-bracket}
\{\Ah,\Bh\}_Q &= -\int_\Omega \frac{\delta\mathcal{A}}{\delta\mathbf{v}} \cdot \left( \Qv \times \dede{\Bh}{\vv} \right) {\rm d}x \\
\label{Ri-v-bracket}
\{\Ah,\Bh\}_{R_i} &= - \int_\Omega q_i \left(\dede{\Ah}{\vv} \cdot  \nabla  \frac{\delta\mathcal{B}}{\delta \rho_i}  - \dede{\Bh}{\vv} \cdot  \nabla  \frac{\delta\mathcal{A}}{\delta \rho_i} \right)  {\rm d}x  \\
\label{S-v-bracket}
\{\Ah,\Bh\}_S &= - \int_\Omega \eta\left(\frac{\delta\mathcal{A}}{\delta \mathbf{v}} \cdot \nabla \frac{\delta\mathcal{B}}{\delta  s} - \frac{\delta\mathcal{B}}{\delta \mathbf{v}} \cdot \nabla \frac{\delta\mathcal{A}}{\delta  s}\right){\rm d}x
\end{align}
with $\Qv = \frac{\nabla \times \vv}{\rho}$.

\paragraph{Equations of motion.}
The functional derivatives \eqref{dHdx-vv} can be substituted into the Poisson brackets \eqref{Q-v-bracket}--\eqref{S-v-bracket} to yield the equations of motion
\begin{align}
\label{v-v-eom}
&\partial_t \vv + \Qv \times \Fv + \sum_i q_i \nabla B_i + \eta \nabla T = 0 \\
\label{rhoi-v-eom}
&\partial_t \rho_i + \nabla \cdot (q_i \Fv) = 0 \textcolor{white}{\sum_i}\\
\label{S-v-eom}
&\partial_t s + \nabla \cdot (\eta \Fv) = 0
\end{align}
which are the same as \eqref{v-eqn-EPv}--\eqref{S-eqn-EPv}. The more common form
\begin{equation}\label{final_equation}
\begin{aligned}
&\partial_t \uv + \nabla \times \uv \times \uv + 2 \boldsymbol{\Omega} \times \uv + \nabla K + \nabla \Phi + \alpha \nabla p = 0 \\
&\partial_t \rho_i + \nabla \cdot (\rho_i \uv) = 0 \\
&\partial_t s + \nabla \cdot (\eta \rho \uv) = 0
\end{aligned}
\end{equation}
is obtained by substituting in the actual values for functional derivatives (\ref{dHdx-vv}) and noting that
\begin{equation}
\sum_i q_i \nabla B_i + \eta \nabla T = \sum_iq_i (\nabla K + \nabla \Phi) + \sum_i q_i \nabla \mu_i + \eta \nabla T = \nabla K + \nabla \Phi + \alpha \nabla p
\end{equation}
since $\sum_i q_i = 1$ and $\sum_i q_i \nabla \mu_i + \eta \nabla T = \alpha \nabla p$ by \eqref{gibbs-duhem}. We have also used $(\nabla \times \vv) \times \uv = (\nabla \times \uv) \times \uv + 2 \bf \Omega \times \uv$ and $\partial_t \vv = \partial_t \uv$.

\subsection{Dissipation Brackets for $(\vv,\rho_i,s)$}
The variational formulation with irreversible processes yields same equations of motion for $\rho_i$ and $s$ as before. The momentum equation takes the form
\begin{equation}
\label{v-addl-terms}
\partial_t \vv + \Qv \times \Fv + \sum_i q_i \nabla B_i + \eta \nabla T - \frac{1}{\rho} \nabla \cdot \sigmafr = 0.
\end{equation}

\paragraph{Single generator.}
Using the chain rule (\ref{chain-rule-vv-rhoi-s}) in (\ref{single-generator-bracket}) gives the single generator bracket in the variables $(\vv,\rho_i,s)$ as 
\begin{equation}\label{vv-rhoi-s-single-generator-bracket}
\begin{aligned}
{[}\mathcal{A},\mathcal{H}{]}=& - \int _\Omega\nabla\left(\frac{1}{\rho} \frac{\delta \mathcal{A}}{\delta\mathbf{v}}\right) : \sigmafr  \Big(\frac{\delta \mathcal{H}}{\delta x}\Big){\rm d}x+ \int _\Omega\nabla \frac{\delta \mathcal{A}}{\delta s} \cdot \mathbf{j}_s \Big(\frac{\delta \mathcal{H}}{\delta x}\Big){\rm d} x\\
&+\int_\Omega \frac{\frac{\delta \mathcal{A}}{\delta s}}{\frac{\delta \mathcal{H}}{\delta s}}\left(\boldsymbol{\sigma} ^{\rm fr} \Big(\frac{\delta \mathcal{H}}{\delta x}\Big)\!: \!\nabla \left(\frac{1}{\rho} \frac{\delta \mathcal{H}}{\delta\mathbf{v}}\right) - \mathbf{j} _s \Big(\frac{\delta \mathcal{H}}{\delta x}\Big)\! \cdot  \!\nabla  \frac{\delta \mathcal{H} }{\delta s } \textcolor{white}{\sum_i}\right.\\
&\hspace{2cm}\left.- \sum_i\mathbf{j} _i \Big(\frac{\delta \mathcal{H}}{\delta x}\Big) \!\cdot \! \nabla  \frac{\delta \mathcal{H} }{\delta \rho _i } - \sum_i j _i  \Big(\frac{\delta \mathcal{H}}{\delta x}\Big)  \frac{\delta \mathcal{H} }{\delta\rho_i} \right){\rm d}x\\
& +\sum_i \int _\Omega\nabla \frac{\delta \mathcal{A}}{\delta \rho_i} \cdot \mathbf{j}_i \Big(\frac{\delta \mathcal{H}}{\delta x}\Big){\rm d} x+\sum_i \int _\Omega \frac{\delta \mathcal{A}}{\delta \rho_i} j_i \Big(\frac{\delta \mathcal{H}}{\delta x}\Big){\rm d} x.
\end{aligned}
\end{equation}
In deriving this, we have used the mass control conditions to simplify the terms arising from $\frac{\delta \mathsf{A}}{\delta \rho_i}$ and $\frac{\delta \mathsf{H}}{\delta \rho_i}$.

\paragraph{Double generator.} Using the chain rule \eqref{chain-rule-vv-rhoi-s} in \eqref{double-generator-bracket} gives the double generator bracket in the variables $(\vv,\rho_i,s)$ as
\begin{equation}
\label{vv-rhoi-s-double-generator-bracket}
\begin{aligned}
&(\mathcal{A},\mathcal{B})= - \int_\Omega \nabla \left(\frac{1}{\rho}\frac{\delta \mathcal{A}}{\delta \mathbf{v}}\right): \sigmafr \frac{\delta \mathcal{B}}{\delta s} {\rm d}x - \int _\Omega\nabla \left(\frac{1}{\rho}\frac{\delta \mathcal{B}}{\delta \mathbf{v}}\right): \sigmafr \frac{\delta \mathcal{A}}{\delta s}{\rm d}x \\
&+ \int_\Omega \frac{1}{\frac{\delta \mathcal{H}}{\delta s}} \left(\sigmafr : \nabla \left(\frac{1}{\rho}\frac{\delta \mathcal{H}}{\delta \mathbf{v}}\right) - \mathbf{j}_s\cdot \nabla \frac{\delta \mathcal{H}}{\delta s}- \sum_i \mathbf{j}_i \cdot \nabla \frac{\delta \mathcal{H}}{\delta \rho_i} - \sum_i  j_i \frac{\delta \mathcal{H}}{\delta \rho_i} \right) \frac{\delta \mathcal{A}}{\delta s} \frac{\delta \mathcal{B}}{\delta s} {\rm d}x\\
&+ \int_\Omega \nabla \frac{\delta \mathcal{A}}{\delta s} \cdot \mathbf{j}_s \frac{\delta \mathcal{B}}{\delta s}{\rm d}x + \int_\Omega \nabla  \frac{\delta \mathcal{B}}{\delta s}\cdot \mathbf{j}_s \frac{\delta \mathcal{A}}{\delta s} {\rm d}x\\
&+ \sum_i \int_\Omega \nabla \frac{\delta \mathcal{A}}{\delta \rho_i} \cdot \mathbf{j}_i \frac{\delta \mathcal{B}}{\delta s}{\rm d}x + \sum_i \int_\Omega \nabla \frac{\delta \mathcal{B}}{\delta\rho_i} \cdot \mathbf{j}_i \frac{\delta \mathcal{A}}{\delta s}{\rm d}x \\
&+  \sum_i \int _\Omega \frac{\delta \mathcal{A}}{\delta \rho_i} j_i \frac{\delta \mathcal{B}}{\delta s}{\rm d}x+\sum_i \int_\Omega \frac{\delta \mathcal{B}}{\delta \rho_i} j_i \frac{\delta \mathcal{A}}{\delta s}{\rm d}x.
\end{aligned}
\end{equation}
In deriving this, we have again used the mass control conditions to simplify the terms arising from $\frac{\delta \mathsf{A}}{\delta\rho_i}$, $\frac{\delta \mathsf{B}}{\delta\rho_i}$ and $\frac{\delta \mathsf{H}}{\delta\rho_i}$.

\subsection{The variables $(\vv,\rho_i,\eta)$}

\paragraph{Chain rule.} If we predict the specific entropy $\eta$ instead of the entropy density $s$, we can write $\Aph[\vv,\rho_i,\eta] = \Ah[\vv,\rho_i,s]$ and the chain rule gives
\begin{eqnarray}
\frac{\delta \mathcal{A}'}{\delta \rho_i} = \dede{\Ah}{\rho_i} + \eta \frac{\delta \mathcal{A}}{\delta s}, \quad\quad
\frac{\delta \mathcal{A}'}{\delta \mathbf{v}}= \frac{\delta \mathcal{A}}{\delta \mathbf{v}}, \quad\quad
\frac{\delta \mathcal{A}'}{\delta \eta} = \rho \frac{\delta \mathcal{A}}{\delta s} .
\end{eqnarray}

\paragraph{Hamiltonian.}
Therefore we have $\Hph[\rho_i,\vv,s] = \Hh[\rho_i,\vv,S]$ and 
\begin{eqnarray}
\label{dHdx-vv-s}
\frac{\delta \mathcal{H}'}{\delta \rho_i} := B^\prime_i = K + \Phi + \mu_i + sT, \quad\quad
\frac{\delta \mathcal{H}'}{\delta \mathbf{v}} := \Fv = \rho \mathbf{u}, \quad\quad
\frac{\delta \mathcal{H}'}{\delta \eta} = \rho T  .
\end{eqnarray}

\paragraph{Poisson Bracket.}
Using also the chain rule, the Poisson bracket \eqref{Q-v-bracket}--\eqref{S-v-bracket} becomes
\begin{align}
\label{Q-s-bracket}
\{\Aph,\Bph\}_Q &= \int_\Omega \Qv \cdot \left( \frac{\delta \mathcal{A}'}{\delta \mathbf{v}} \times \frac{\delta \mathcal{B}'}{\delta \mathbf{v}} \right) {\rm d}x \\
\label{Ri-s-bracket}
\{\Aph,\Bph\}_{R_i} &= \int_\Omega \left(-\frac{\delta \mathcal{A}'}{\delta \rho_i} \nabla \cdot \left(q_i \frac{\delta \mathcal{B}'}{\delta \mathbf{v}}\right) + \frac{\delta \mathcal{B}'}{\delta \rho_i} \nabla \cdot \left(q_i \frac{\delta \mathcal{A}'}{\delta \mathbf{v}}\right)\right){\rm d}x\\
\label{s-s-bracket}
\{\Aph,\Bph\}_s  &= \int_\Omega \frac{\nabla \eta}{\rho} \cdot \left(\frac{\delta \mathcal{A}'}{\delta \mathbf{v}} \frac{\delta \mathcal{B}'}{\delta \eta} - \frac{\delta \mathcal{B}'}{\delta \mathbf{v}}  \frac{\delta \mathcal{A}'}{\delta \eta}\right) {\rm d}x.
\end{align}

\paragraph{Single generator bracket.} The single generator dissipation bracket \eqref{single-generator-bracket} in the variables $(\vv,\rho_i,\eta)$ becomes
\begin{equation}\label{vv-rhoi-eta-single-generator-bracket}
\begin{aligned}
{[}\mathcal{A}',\mathcal{H}'{]}=& - \int _\Omega\nabla\left(\frac{1}{\rho} \frac{\delta \mathcal{A}'}{\delta\mathbf{v}}\right) : \sigmafr  \Big(\frac{\delta \mathcal{H}'}{\delta x}\Big){\rm d}x+ \int _\Omega\nabla \left(\frac{1}{\rho}\frac{\delta \mathcal{A}'}{\delta \eta}\right) \cdot \mathbf{j}_s \Big(\frac{\delta \mathcal{H}'}{\delta x}\Big){\rm d} x\\
&+\int_\Omega \frac{\frac{\delta \mathsf{A}'}{\delta \eta}}{\frac{\delta \mathcal{H}'}{\delta \eta}}\left(\boldsymbol{\sigma} ^{\rm fr} \Big(\frac{\delta \mathcal{H}'}{\delta x}\Big)\!: \!\nabla \left(\frac{1}{\rho} \frac{\delta \mathcal{A}'}{\delta\mathbf{v}}\right) - \mathbf{j} _s \Big(\frac{\delta \mathcal{H}'}{\delta x}\Big)\! \cdot  \!\nabla \left(\frac{1}{\rho} \frac{\delta \mathcal{H}'}{\delta \eta }\right) \textcolor{white}{\sum_i}\right.\\
&\hspace{2cm}\left.- \sum_i\mathbf{j} _i \Big(\frac{\delta \mathcal{H}'}{\delta x}\Big) \!\cdot \! \nabla  \frac{\delta \mathcal{H}'}{\delta \rho _i } - \sum_i j _i  \Big(\frac{\delta \mathcal{H}'}{\delta x}\Big)  \frac{\delta \mathcal{H}'}{\delta\rho_i} \right){\rm d}x\\
& +\sum_i \int _\Omega\nabla \frac{\delta \mathcal{A}'}{\delta \rho_i} \cdot \mathbf{j}_i \Big(\frac{\delta \mathcal{H}'}{\delta x}\Big){\rm d} x+\sum_i \int _\Omega \frac{\delta \mathcal{A}'}{\delta \rho_i} j_i \Big(\frac{\delta \mathcal{H}'}{\delta x}\Big){\rm d} x,
\end{aligned}
\end{equation}
where we have used the mass control conditions to simplify the terms arising from $\frac{\delta \mathcal{A}}{\delta \rho_i}$ and $\frac{\delta \mathcal{H}}{\delta\rho_i}$.

\paragraph{Double generator bracket.} The double generator dissipation bracket \eqref{double-generator-bracket} in the variables $(\vv,\rho_i,\eta)$ is
\begin{equation}
\label{vv-rhoi-eta-double-generator-bracket}
\begin{aligned}
&(\mathcal{A}',\mathcal{B}')= - \int_\Omega \nabla \left(\frac{1}{\rho}\frac{\delta \mathcal{A}'}{\delta \mathbf{v}}\right): \sigmafr  \frac{1}{\rho}\frac{\delta \mathcal{B}'}{\delta \eta} {\rm d}x - \int _\Omega\nabla \left(\frac{1}{\rho}\frac{\delta \mathcal{B}'}{\delta \mathbf{v}}\right): \sigmafr  \frac{1}{\rho}\frac{\delta \mathcal{A}'}{\delta \eta} {\rm d}x \\
&+ \int_\Omega \frac{1}{\frac{\delta \mathcal{H}'}{\delta \eta}} \left(\sigmafr : \nabla \left(\frac{1}{\rho}\frac{\delta \mathcal{H}'}{\delta \mathbf{v}}\right) - \mathbf{j}_s\cdot \nabla  \left(\frac{1}{\rho}\frac{\delta \mathcal{H}'}{\delta \eta}\right)- \sum_i \mathbf{j}_i \cdot \nabla \frac{\delta \mathcal{H}'}{\delta \rho_i} - \sum_i  j_i \frac{\delta \mathcal{H}'}{\delta \rho_i} \right) \frac{1}{\rho}\frac{\delta \mathcal{A}'}{\delta \eta} \frac{\delta \mathcal{B}'}{\delta \eta} {\rm d}x\\
&+ \int_\Omega \nabla  \left(\frac{1}{\rho}\frac{\delta \mathcal{A}'}{\delta \eta}\right) \cdot \mathbf{j}_s \frac{1}{\rho}\frac{\delta \mathcal{B}'}{\delta \eta}{\rm d}x + \int_\Omega \nabla  \left(\frac{1}{\rho} \frac{\delta \mathcal{B}'}{\delta \eta}\right)\cdot \mathbf{j}_s \frac{1}{\rho}\frac{\delta \mathcal{A}'}{\delta \eta} {\rm d}x\\
&+ \sum_i \int_\Omega \nabla \frac{\delta \mathcal{A}'}{\delta \rho_i} \cdot \mathbf{j}_i \frac{1}{\rho}\frac{\delta \mathcal{B}'}{\delta \eta}{\rm d}x + \sum_i \int_\Omega \nabla \frac{\delta \mathcal{B}'}{\delta\rho_i} \cdot \mathbf{j}_i \frac{1}{\rho}\frac{\delta \mathcal{A}'}{\delta \eta}{\rm d}x \\
&+  \sum_i \int _\Omega \frac{\delta \mathcal{A}'}{\delta \rho_i} j_i \frac{1}{\rho}\frac{\delta \mathcal{B}'}{\delta \eta}{\rm d}x+\sum_i \int_\Omega \frac{\delta \mathcal{B}'}{\delta \rho_i} j_i \frac{1}{\rho}\frac{\delta \mathcal{A}'}{\delta \eta}{\rm d}x,
\end{aligned}
\end{equation}
where we have used the mass control conditions to simplify the terms arising from $\frac{\delta \mathcal{A}}{\delta \rho_i}$, $\frac{\delta \mathcal{B}}{\delta \rho_i}$ and $\frac{\delta \mathcal{H}}{\delta \rho_i}$.

\paragraph{Equations of motion.} Using the functional derivatives \eqref{dHdx-vv-s} in the Poisson brackets \eqref{Q-s-bracket}--\eqref{s-s-bracket} and either of the dissipation brackets \eqref{vv-rhoi-eta-single-generator-bracket} or \eqref{vv-rhoi-eta-double-generator-bracket} gives the equations of motion as
\begin{align*}
&\partial_t \rho_i + \nabla \cdot (q_i \Fv) + \nabla \cdot \mathbf{j}_i - j_i = 0 \\
&\partial_t \vv + \Qv \times \Fv + \sum_i q_i \nabla B^\prime_i - T \nabla \eta - \frac{1}{\rho} \nabla \cdot \sigmafr = 0 \\
&\partial_t \eta + \frac{1}{\rho} \mathbf{F}\cdot \nabla \eta + \frac{q}{\rho} \nabla \cdot \mathbf{j}_s - \frac{1}{\rho T} \sigmafr : \nabla \uv + \frac{1}{\rho T} \js \cdot \nabla T + \frac{1}{\rho T} \sum_i \left( \mathbf{j}_i\cdot \nabla \mu_i + j_i \mu_i\right) = 0 .
\end{align*}

\subsection{The variables $(\vv,\rho,q_k,s)$}

\paragraph{Chain rule.} Here, besides $\mathbf{v}$ and $s$, the variables are the total density $\rho$ plus $n-1$ sparse concentrations $q_k$, where $k\in \{1,\dots,n-1\}$. We can write $\mathscr{A}[\vv,\rho,q_k,s] = \mathcal{A}[\vv,\rho_i,s]$ and the chain rule gives
\begin{equation}\label{change_variable_A4}
\begin{aligned}
\frac{\delta \mathcal{A}}{\delta \mathbf{v}}&=\frac{\delta \mathscr{A}}{\delta \mathbf{v}}, \quad\quad
\frac{\delta \mathcal{A}}{\delta \rho_d}= \frac{\delta \mathscr{A}}{\delta \rho}- \frac{1}{\rho}\sum_{k} \frac{\delta \mathscr{A}}{\delta q_{k}}q_{k}\\
\frac{\delta \mathcal{A}}{\delta \rho_k} &= \frac{\delta \mathscr{A}}{\delta\rho}  - \frac{1}{\rho}\sum_{k'} \frac{\delta \mathscr{A}}{\delta q_{k'}}q_{k'}+ \frac{1}{\rho} \frac{\delta \mathscr{A}}{\delta q_k}, \quad\quad
  \frac{\delta \mathcal{A}}{\delta s}=\frac{\delta \mathscr{A}}{\delta s}.
\end{aligned}
\end{equation}
where $k' \in \{1,\dots,n-1\}$ is the set of sparse components.

\paragraph{Hamiltonian.} The Hamiltonian $\mathscr{H}[\vv,\rho,q_k,s]$ is given as
\begin{equation}
\mathscr{H}[\vv,\rho,q_k,s]= \int_\Omega \rho \left[ K + \Phi + U \right] {\rm d}x,
\end{equation}
where $U(\alpha,\eta,q_k,\chi_d) = U(\frac{1}{\rho},\frac{s}{\rho},q_k,1 - \sum_k q_k)$ with $\chi_d$ the concentration of the dominant component. The functional derivatives of $\mathcal{H}$ are given by
\begin{equation}
\begin{aligned}
\label{dHdx-rho}
&\frac{\delta\mathscr{H}}{\delta \rho} := B = K + \Phi + U + p\alpha - \eta T, \quad\quad\frac{\delta\mathscr{H}}{\delta q_k} := \rho (\mu_k - \mu_d) \\
&\frac{\delta\mathscr{H}}{\delta \vv} := \Fv = \rho \mathbf{u}, \quad\quad\frac{\delta\mathscr{H}}{\delta  s} := T.
\end{aligned}
\end{equation}

\paragraph{Poisson bracket.} The Poisson bracket in the variables $(\vv,\rho,q_k,s)$ can again be expressed as the sum of $n+2$ brackets
\begin{equation}
\{\mathscr{A},\mathscr{B}\} = \{\mathscr{A},\mathscr{B}\}_Q  + \{\mathscr{A},\mathscr{B}\}_R + \sum_k \{\mathscr{A},\mathscr{B}\}_{q_k} + \{\mathscr{A},\mathscr{B}\}_S
\end{equation}
with
\begin{align}
\label{Q-chi-bracket}
\{\mathscr{A},\mathscr{B}\}_Q &= \int_\Omega \Qv \cdot \left( \frac{\delta\mathscr{A}}{\delta \mathbf{v}}\times \frac{\delta\mathscr{B}}{\delta \mathbf{v}}\right) {\rm d}x\\
\label{R-chi-bracket}
\{\mathscr{A},\mathscr{B}\}_{R} &= \int_\Omega \left(- \frac{\delta\mathscr{A}}{\delta \rho} \nabla \cdot \frac{\delta\mathscr{B}}{\delta \mathbf{v}} + \frac{\delta\mathscr{B}}{\delta \rho} \nabla \cdot\frac{\delta\mathscr{A}}{\delta \mathbf{v}} \right){\rm d}x \\
\label{qk-chi-bracket}
\{\mathscr{A},\mathscr{B}\}_{q_k} &= \int_\Omega \frac{\nabla q_k}{\rho} \cdot \left(\frac{\delta\mathscr{A}}{\delta \mathbf{v}} \frac{\delta\mathscr{B}}{\delta q_k} - \dede{\mathscr{B}}{\vv} \dede{\mathscr{A}}{q_k}\right) {\rm d}x \\
\label{S-chi-bracket}
\{\mathscr{A},\mathscr{B}\}_S &=\int_\Omega \left(- \frac{\delta\mathscr{A}}{\delta s} \nabla \cdot \left( \eta \frac{\delta\mathscr{B}}{\delta \mathbf{v}} \right) + \frac{\delta\mathscr{B}}{\delta s} \nabla \cdot \left( \eta \frac{\delta\mathscr{A}}{\delta \mathbf{v}}\right) \right){\rm d}x.
\end{align}

\paragraph{Single generator bracket.} Using \eqref{change_variable_A4} in \eqref{vv-rhoi-s-single-generator-bracket}, 
the single generator dissipation bracket in the variables $(\vv,\rho,q_k,s)$ becomes
\begin{equation}\label{vv-rho-qk-s-single-generator-bracket}
\begin{aligned}
{[}\mathscr{A},\mathscr{H}{]}=& - \int _\Omega\nabla\left(\frac{1}{\rho} \frac{\delta \mathscr{A}}{\delta\mathbf{v}}\right) : \sigmafr  \Big(\frac{\delta \mathscr{H}}{\delta x}\Big){\rm d}x+ \int _\Omega\nabla \frac{\delta \mathscr{A}}{\delta s}\cdot \mathbf{j}_s \Big(\frac{\delta \mathscr{H}}{\delta x}\Big){\rm d} x\\
&+\int_\Omega \frac{\frac{\delta \mathscr{A}}{\delta s}}{\frac{\delta \mathscr{H}}{\delta s}}\left(\boldsymbol{\sigma} ^{\rm fr} \Big(\frac{\delta \mathscr{H}}{\delta x}\Big)\!: \!\nabla \left(\frac{1}{\rho} \frac{\delta \mathscr{H}}{\delta\mathbf{v}}\right) - \mathbf{j} _s \Big(\frac{\delta \mathscr{H}}{\delta x}\Big)\! \cdot  \!\nabla   \frac{\delta \mathscr{H}}{\delta s }  \textcolor{white}{\sum_i}\right.\\
&\hspace{2cm}\left.- \sum_k\mathbf{j} _k \Big(\frac{\delta \mathscr{H}}{\delta x}\Big) \!\cdot \! \nabla \left(\frac{1}{\rho} \frac{\delta \mathscr{H}}{\delta q_k}\right) - \sum_k j _k  \Big(\frac{\delta \mathscr{H}}{\delta x}\Big)  \frac{1}{\rho}\frac{\delta \mathscr{H}}{\delta q_k} \right){\rm d}x\\
& +\sum_k \int _\Omega\nabla \left(\frac{1}{\rho} \frac{\delta \mathscr{A}}{\delta q_k} \right)\cdot \mathbf{j}_k \Big(\frac{\delta \mathscr{H}}{\delta x}\Big){\rm d} x+\sum_k \int _\Omega \frac{1}{\rho}\frac{\delta \mathscr{A}}{\delta q_k} j_k \Big(\frac{\delta \mathscr{H}}{\delta x}\Big){\rm d} x.
\end{aligned}
\end{equation}

\paragraph{Double generator bracket.} Using \eqref{change_variable_A4} in \eqref{vv-rhoi-s-double-generator-bracket}, the double generator dissipation bracket becomes
\begin{equation}
\label{vv-rho-qk-s-double-generator-bracket}
\begin{aligned}
&(\mathscr{A},\mathscr{B})= - \int_\Omega \nabla \left(\frac{1}{\rho}\frac{\delta \mathscr{A}}{\delta \mathbf{v}}\right): \sigmafr \frac{\delta \mathscr{B}}{\delta s} {\rm d}x - \int _\Omega\nabla \left(\frac{1}{\rho}\frac{\delta \mathscr{B}}{\delta \mathbf{v}}\right): \sigmafr \frac{\delta \mathscr{A}}{\delta s}{\rm d}x \\
&+ \int_\Omega \frac{1}{\frac{\delta \mathscr{H}}{\delta s}} \left(\sigmafr : \nabla \left(\frac{1}{\rho}\frac{\delta \mathscr{H}}{\delta \mathbf{v}}\right) - \mathbf{j}_s\cdot \nabla \frac{\delta \mathscr{H}}{\delta s}- \sum_k \mathbf{j}_k \cdot \nabla \left(\frac{1}{\rho}\frac{\delta \mathscr{H}}{\delta q_k}\right) - \sum_k  j_k \frac{1}{\rho}\frac{\delta \mathscr{H}}{\delta q_k} \right) \frac{\delta \mathscr{A}}{\delta s} \frac{\delta \mathscr{B}}{\delta s} {\rm d}x\\
&+ \int_\Omega \nabla \frac{\delta \mathscr{A}}{\delta s} \cdot \mathbf{j}_s \frac{\delta \mathscr{B}}{\delta s}{\rm d}x + \int_\Omega \nabla  \frac{\delta \mathscr{B}}{\delta s}\cdot \mathbf{j}_s \frac{\delta \mathscr{A}}{\delta s} {\rm d}x\\
&+ \sum_k \int_\Omega \nabla\left(\frac{1}{\rho} \frac{\delta \mathscr{A}}{\delta q_k} \right)\cdot \mathbf{j}_k \frac{\delta \mathscr{B}}{\delta s}{\rm d}x + \sum_k \int_\Omega \nabla \left(\frac{1}{\rho}\frac{\delta \mathscr{B}}{\delta q_k} \right)\cdot \mathbf{j}_k \frac{\delta \mathscr{A}}{\delta s}{\rm d}x \\
&+  \sum_k \int _\Omega \frac{1}{\rho}\frac{\delta \mathscr{A}}{\delta q_k} j_k \frac{\delta \mathscr{B}}{\delta s}{\rm d}x+\sum_k \int_\Omega \frac{1}{\rho}\frac{\delta \mathscr{B}}{\delta q_k} j_k \frac{\delta \mathscr{A}}{\delta s}{\rm d}x.
\end{aligned}
\end{equation}

\paragraph{Entropy generation.} Both dissipation brackets give the entropy generation rate as
\begin{align*}
&[\mathscr{S},\mathscr{H}] = (\mathscr{S},\mathscr{S})\\
&= \int_\Omega \frac{1}{\frac{\delta \mathscr{H}}{\delta s}} \left( \sigmafr : \nabla \left( \frac{1}{\rho} \frac{\delta \mathscr{H}}{\delta \mathbf{v}}\right) - \mathbf{j}_s \cdot \nabla \frac{\delta \mathscr{H}}{\delta s} - \sum_k \mathbf{j}_k \cdot \nabla \left(\frac{1}{\rho} \frac{\delta \mathscr{H}}{\delta q_k}\right) - \sum_k j_k\frac{1}{\rho} \frac{\delta \mathscr{H}}{\delta q_k} \right){\rm d}x.
\end{align*}
Plugging in the actual values for $\frac{\delta \mathscr{H}}{\delta x}$, this is
\begin{equation}
\int _\Omega\frac{1}{T} \left( \sigmafr : \nabla \uv - \mathbf{j}_s \cdot \nabla T - \sum_k \mathbf{j}_k \cdot \nabla \left(\mu_k - \mu_d\right) - \sum_k j_k \left(\mu_k - \mu_d \right) \right){\rm d}x.
\end{equation}
The last two terms can be rewritten using $\sum_k \mathbf{j}_k = - \mathbf{j}_d$ and $\sum_k j_k = -j_i$ to finally yield
\begin{equation}
\int_ \Omega\frac{1}{T} \left( \sigmafr : \nabla \uv - \mathbf{j}_s \cdot \nabla T - \sum_i \mathbf{j}_i \cdot \nabla \mu_i - \sum_i j_i \mu_i \right){\rm d}x
\end{equation}
as expected. Therefore we see that the same parameterizations can be used as before, since the form of the entropy generation is identical.

\paragraph{Equations of motion.} Using the functional derivatives \eqref{dHdx-rho} in the Poisson brackets \eqref{Q-chi-bracket}--\eqref{S-chi-bracket} and either of the dissipation bracket \eqref{vv-rho-qk-s-single-generator-bracket} or \eqref{vv-rho-qk-s-double-generator-bracket}, the equations of motion are
\begin{align*}
&\partial_t \rho + \nabla \cdot \Fv = 0 \\
&\partial_t q_k + \uv \cdot \nabla q_k + \frac{1}{\rho} \nabla \cdot \mathbf{j}_k - \frac{1}{\rho}j_k = 0 \\
&\partial_t \vv + \Qv \times \Fv + \nabla B + \eta T - \sum_k (\mu_k - \mu_d) \nabla q_k - \frac{1}{\rho} \nabla \cdot \sigmafr = 0 \\
&\partial_t s + \nabla \cdot (\eta \Fv) + \nabla \cdot \js - \frac{1}{T} \sigmafr : \nabla \uv + \frac{1}{T} \js \cdot \nabla T + \frac{1}{T} \sum_i \left( \ji\cdot \nabla \mu_i + j_i \mu_i\right) = 0.
\end{align*}
Note the sum is over $i$ in the $s$ equation, while the sum is over $k$ in the $\vv$ equation. However, the term $\sum_k (\mu_k - \mu_d) \nabla q_k$ is equal to $\sum_i \mu_i \nabla q_i$, as before, since $\sum_k q_k = 1 - q_d$.

\subsection{The variables $(\vv,\rho,q_k,\eta)$}
\label{alt-bannon-prongostic}

When the variables $(\vv,\rho,q_k,\eta)$ are chosen, the resulting bracket formulation extends the formulation of \cite{Bannon2003} to additional components, and with irreversible processes.

\paragraph{Chain rule.} Again, if we predict the specific entropy $\eta$ instead of the entropy density $s$, we can write $\mathscr{A}'[\vv,\rho,q_k,\eta] = \mathscr{A}[\vv,\rho,q_k,s]$ and the chain rule gives
\begin{eqnarray}
\label{chain-rule-rho-qk}
\frac{\delta \mathscr{A}'}{\delta \rho} = \frac{\delta \mathscr{A}}{\delta \rho} + \eta \frac{\delta \mathscr{A}}{\delta s} \quad\quad
\frac{\delta \mathscr{A}'}{\delta q_k} = \frac{\delta \mathscr{A}}{\delta q_k} \quad\quad
\frac{\delta \mathscr{A}'}{\delta \mathbf{v}} = \frac{\delta \mathscr{A}}{\delta \mathbf{v}} \quad\quad
\frac{\delta \mathscr{A}'}{\delta \eta} = \rho \frac{\delta \mathscr{A}}{\delta s}.
\end{eqnarray}

\paragraph{Hamiltonian.} Therefore we have $\mathscr{H}'[\vv,\rho,q_k,\eta] = \mathscr{H}[\vv,\rho,q_k,s]$ and 
\begin{equation}\label{dHdx-rho-eta}
\begin{aligned}
&\frac{\delta \mathscr{H}'}{\delta \rho}  := B^\prime = K + \Phi + U + p \alpha, \quad\quad
\frac{\delta \mathscr{H}'}{\delta q_k}  := \rho (\mu_k - \mu_d),\\
&\frac{\delta \mathscr{H}'}{\delta \mathbf{v}}  := \Fv = \rho \uv, \quad\quad
\frac{\delta \mathscr{H}'}{\delta  \eta} = \rho T .
\end{aligned}
\end{equation}

\paragraph{Poisson bracket.}
Using also the chain rule \eqref{chain-rule-rho-qk}, the Poisson brackets \eqref{Q-chi-bracket}--\eqref{S-chi-bracket} become
\begin{eqnarray}
\label{Q-s-chi-bracket}
\{\mathscr{A}',\mathscr{B}'\}_Q &=& \int_\Omega \Qv \cdot \left( \frac{\delta \mathscr{A}'}{\delta \mathbf{v}}\times \frac{\delta \mathscr{B}'}{\delta \mathbf{v}}\right) {\rm d}x\\
\label{R-s-chi-bracket}
\{\mathscr{A}',\mathscr{B}'\}_{R} &=& \int_\Omega \left(-\frac{\delta \mathscr{A}'}{\delta \rho} \nabla \cdot \frac{\delta \mathscr{B}'}{\delta \mathbf{v}} + \frac{\delta \mathscr{B}'}{\delta \rho} \nabla \cdot \frac{\delta \mathscr{A}'}{\delta \mathbf{v}}\right) {\rm d}x\\
\label{qk-s-chi-bracket}
\{\mathscr{A}',\mathscr{B}'\}_{q_k} &=& \int_\Omega \frac{\nabla q_k}{\rho} \cdot \left(\frac{\delta \mathscr{A}'}{\delta \mathbf{v}} \frac{\delta \mathscr{B}'}{\delta q_k} - \frac{\delta \mathscr{B}'}{\delta \mathbf{v}} \frac{\delta \mathscr{A}'}{\delta q_k}\right) {\rm d}x \\
\label{s-s-chi-bracket}
\{\mathscr{A}',\mathscr{B}'\}_S  &=& \int_\Omega \frac{\nabla \eta}{\rho} \cdot \left(\frac{\delta \mathscr{A}'}{\delta \mathbf{v}} \frac{\delta \mathscr{B}'}{\delta \eta} - \frac{\delta \mathscr{B}'}{\delta \mathbf{v}} \frac{\delta \mathscr{A}'}{\delta \eta}\right){\rm d}x. 
\end{eqnarray}

\paragraph{Single generator bracket.}
The single generator dissipation bracket is

\begin{equation}\label{vv-rho-qk-eta-single-generator-bracket}
\begin{aligned}
{[}\mathscr{A}',\mathscr{H}'{]}=& - \int _\Omega\nabla\left(\frac{1}{\rho} \frac{\delta \mathscr{A}'}{\delta\mathbf{v}}\right) : \sigmafr  \Big(\frac{\delta \mathscr{H}'}{\delta x}\Big){\rm d}x+ \int _\Omega\nabla \left(\frac{1}{\rho}\frac{\delta \mathscr{A}'}{\delta \eta}\right)\cdot \mathbf{j}_s \Big(\frac{\delta \mathscr{H}'}{\delta x}\Big){\rm d} x\\
&+\int_\Omega \frac{\frac{\delta \mathscr{A}'}{\delta \eta}}{\frac{\delta \mathscr{H}'}{\delta \eta}}\left(\boldsymbol{\sigma} ^{\rm fr} \Big(\frac{\delta \mathscr{H}'}{\delta x}\Big)\!: \!\nabla \left(\frac{1}{\rho} \frac{\delta \mathscr{H}'}{\delta\mathbf{v}}\right) - \mathbf{j} _s \Big(\frac{\delta \mathscr{H}'}{\delta x}\Big)\! \cdot  \!\nabla  \left(\frac{1}{\rho} \frac{\delta \mathscr{H}'}{\delta \eta }\right)  \textcolor{white}{\sum_i}\right.\\
&\hspace{2cm}\left.- \sum_k\mathbf{j} _k \Big(\frac{\delta \mathscr{H}'}{\delta x}\Big) \!\cdot \! \nabla \left(\frac{1}{\rho} \frac{\delta \mathscr{H}}{\delta q_k}\right) - \sum_k j _k  \Big(\frac{\delta \mathscr{H}'}{\delta x}\Big)  \frac{1}{\rho}\frac{\delta \mathscr{H}'}{\delta q_k} \right){\rm d}x\\
& +\sum_k \int _\Omega\nabla \left(\frac{1}{\rho} \frac{\delta \mathscr{A}'}{\delta q_k} \right)\cdot \mathbf{j}_k \Big(\frac{\delta \mathscr{H}'}{\delta x}\Big){\rm d} x+\sum_k \int _\Omega \frac{1}{\rho}\frac{\delta \mathscr{A}'}{\delta q_k} j_k \Big(\frac{\delta \mathscr{H}'}{\delta x}\Big){\rm d} x.
\end{aligned}
\end{equation}

\paragraph{Double generator bracket.}
The double generator dissipation bracket is

\begin{equation}
\label{vv-rho-qk-eta-double-generator-bracket}
\begin{aligned}
&(\mathscr{A}',\mathscr{B}')= - \int_\Omega \nabla \left(\frac{1}{\rho}\frac{\delta \mathscr{A}'}{\delta \mathbf{v}}\right): \sigmafr \frac{1}{\rho}\frac{\delta \mathscr{B}'}{\delta \eta} {\rm d}x - \int _\Omega\nabla \left(\frac{1}{\rho}\frac{\delta \mathscr{B}'}{\delta \mathbf{v}}\right): \sigmafr \frac{1}{\rho}\frac{\delta \mathscr{A}'}{\delta \eta}{\rm d}x \\
&+ \int_\Omega \frac{1}{\frac{\delta \mathscr{H}'}{\delta \eta}} \left(\sigmafr : \nabla \left(\frac{1}{\rho}\frac{\delta \mathscr{H}'}{\delta \mathbf{v}}\right) - \mathbf{j}_s\cdot \nabla \left(\frac{1}{\rho}\frac{\delta \mathscr{H}'}{\delta \eta}\right)- \sum_k \mathbf{j}_k \cdot \nabla \left(\frac{1}{\rho}\frac{\delta \mathscr{H}'}{\delta q_k}\right) - \sum_k  j_k \frac{1}{\rho}\frac{\delta \mathscr{H}'}{\delta q_k} \right) \frac{1}{\rho}\frac{\delta \mathscr{A}'}{\delta \eta} \frac{\delta \mathscr{B}'}{\delta \eta} {\rm d}x\\
&+ \int_\Omega \nabla \frac{\delta \mathscr{A}'}{\delta s} \cdot \mathbf{j}_s \frac{\delta \mathscr{B}'}{\delta s}{\rm d}x + \int_\Omega \nabla  \left(\frac{1}{\rho}\frac{\delta \mathscr{B}'}{\delta \eta}\right)\cdot \mathbf{j}_s \frac{1}{\rho}\frac{\delta \mathscr{A}'}{\delta \eta} {\rm d}x\\
&+ \sum_k \int_\Omega \nabla\left(\frac{1}{\rho} \frac{\delta \mathscr{A}'}{\delta q_k} \right)\cdot \mathbf{j}_k \frac{1}{\rho}\frac{\delta \mathscr{B}'}{\delta \eta}{\rm d}x + \sum_k \int_\Omega \nabla \left(\frac{1}{\rho}\frac{\delta \mathscr{B}'}{\delta q_k} \right)\cdot \mathbf{j}_k \frac{1}{\rho}\frac{\delta \mathscr{A}'}{\delta \eta}{\rm d}x \\
&+  \sum_k \int _\Omega \frac{1}{\rho}\frac{\delta \mathscr{A}'}{\delta q_k} j_k \frac{1}{\rho}\frac{\delta \mathscr{B}'}{\delta \eta}{\rm d}x+\sum_k \int_\Omega \frac{1}{\rho}\frac{\delta \mathscr{B}'}{\delta q_k} j_k \frac{1}{\rho}\frac{\delta \mathscr{A}'}{\delta \eta}{\rm d}x.
\end{aligned}
\end{equation}

\paragraph{Equations of motion.} Using the functional derivatives \eqref{dHdx-rho-eta} in the Poisson brackets \eqref{Q-s-chi-bracket}--\eqref{s-s-chi-bracket} and either of the dissipation brackets \eqref{vv-rho-qk-eta-single-generator-bracket} or \eqref{vv-rho-qk-eta-double-generator-bracket}, the equations of motion are
\begin{align*}
&\partial_t \rho + \nabla \cdot \Fv = 0 \\
&\partial_t q_k + \uv \cdot \nabla q_k + \frac{1}{\rho} \nabla \cdot \mathbf{j}_k - \frac{1}{\rho} j_k= 0 \\
&\partial_t \vv + \Qv \times \Fv + \nabla B^\prime - T \nabla \eta - \sum_k (\mu_k - \mu_d) \nabla q_k - \frac{1}{\rho} \nabla \cdot \sigmafr = 0\\
&\partial_t \eta + \frac{1}{\rho} \mathbf{F}\cdot \nabla \eta + \frac{1}{\rho} \nabla \cdot \mathbf{j}_s- \frac{1}{\rho T} \sigmafr : \nabla \uv + \frac{1}{\rho T} \mathbf{j}_s \cdot \nabla T + \frac{1}{\rho T} \sum_i \left( \mathbf{j}_i\cdot \nabla \mu_i + j_i \mu_i\right) = 0 .
\end{align*}
Note the sum is over $i$ in the $\eta$ equation, while the sum is over $k$ in the $\vv$ equation.

\section{Potential Vorticity and Kelvin Circulation Theorem}
\label{pv-kelvin-circulation}
\subsection{Kelvin Circulation Theorem}
\label{kelvin-circ-theorem}
Integration of (\ref{EP-v}) and (\ref{v-addl-terms}) around a closed loop $\gamma(t)$ moving with the fluid gives the Kelvin circulation theorem
\begin{equation}
\frac{d}{dt} \oint_{\gamma(t)} \vv \cdot \mathbf{dx} = - \oint_{\gamma(t)} \left( \sum_i q_i \nabla \frac{\delta\mathcal{L}}{\delta \rho_i} + \eta \nabla\frac{\delta\mathcal{L}}{\delta s}  - \frac{1}{\rho} \nabla \cdot \sigmafr \right) \cdot \mathbf{dx},
\end{equation}
see \cite{GayBalmaz2017}. Using \eqref{bi-fv-defs}, this can be written as
\begin{equation}
\frac{d}{dt} \oint_{\gamma(t)} \vv \cdot \mathbf{dx} = - \oint_{\gamma(t)} \left( \sum_i q_i \nabla B_i + \eta \nabla T - \frac{1}{\rho} \nabla \cdot \sigmafr \right) \cdot \mathbf{dx} .
\end{equation}

\paragraph{Specific Lagrangian.}
For the specific Lagrangian \eqref{specific-lagrangian} we have $B_i = K + \Phi + \mu_i$ and therefore (using $\sum_i q_i = 1$)
\begin{equation}
\frac{d}{dt} \oint_{\gamma(t)} \vv \cdot \mathbf{dx} = - \oint_{\gamma(t)} \left( \sum_i q_i \nabla \mu_i + \eta \nabla T - \frac{1}{\rho} \nabla \cdot \sigmafr \right) \cdot \mathbf{dx} .
\end{equation}
Using Gibbs-Duhem relationship \eqref{gibbs-duhem}, this can be put into a final simplified form
\begin{equation}
\frac{d}{dt} \oint_{\gamma(t)} \vv \cdot \mathbf{dx} = - \oint_{\gamma(t)} \left( \alpha \nabla p - \frac{1}{\rho} \nabla \cdot \sigmafr \right) \cdot \mathbf{dx} .
\end{equation}

\subsection{Potential Vorticity}
A generalized potential vorticity is defined as
\begin{equation}
q = \frac{\boldsymbol{\eta} \cdot \nabla \psi}{\rho}
\end{equation}
where $\boldsymbol{\eta} = \nabla \times \vv$ and $\psi$ is any scalar function satisfing $D_t \psi = \dot{\psi}$ (for example, $\eta$), with $D_t$ the material derivative. From the system \eqref{system_Eulerian_atmosphere_moist_L}, we get the evolution equations for $q$ and mass-weighted generalized potential vorticity $\rho q$
\begin{equation}
\partial_t q + \uv \cdot \nabla q + \frac{1}{\rho} \nabla \cdot \left( \left[ \sum_i q_i \nabla \mu_i + \eta \nabla T - \frac{1}{\rho} \nabla \cdot \sigmafr \right] \times \nabla \psi + \dot{\psi} \boldsymbol{\eta} \right) = 0
\end{equation}
\begin{equation}
\label{mass-pv-eqn}
\partial_t (\rho q) + \nabla \cdot (\uv \rho q) - \nabla \cdot \left( \left[ \sum_i q_i \nabla \mu_i + \eta \nabla T - \frac{1}{\rho} \nabla \cdot \sigmafr \right] \times \nabla \psi + \dot{\psi} \boldsymbol{\eta} \right) = 0.
\end{equation}
Using the Gibbs-Duhem relationship \eqref{gibbs-duhem}, the term $\frac{1}{\rho} \nabla \cdot \left( \left[ \sum_i q_i \nabla \mu_i + \eta \nabla T \right] \times \nabla \psi \right)$ can be rewritten as $-\frac{1}{\rho^3} \nabla \rho \times \nabla p \cdot \psi$.
When $\psi = \psi(q_i,\eta)$ then $D_t \psi = 0$ when there are no irreversible processes. In particular, for a single component fluid, $\psi=\psi(\eta)$ and the term $\nabla \rho \times \nabla p \cdot \nabla\psi$ vanishes since $p=p(\rho, \eta)$, so $D_tq=0$ and potential vorticity is materially conserved. This explains why $\mathsf{C}=\int \rho\Phi(q,\eta){\rm d}x$ is a Casimir for the single component fluid. For a multicomponent fluid $p=p(\rho, q_i,\eta)$ so $\nabla \rho \times \nabla p \cdot \nabla\psi$ does not vanish for any choice of $\psi=\psi(q_i,\eta)$. In particular, $\int_\Omega \rho q{\rm d}x$ is not a Casimir for multicomponent fluids. (\ref{mass-pv-eqn}) also shows the generation of potential vorticity by boundary processes. More discussion of this can be found in \cite{Charron2018a,Charron2018,ScHeGa2003}.

\end{document}